\documentclass[aps,preprint,showpacs,floatfix,nobibnotes,superscriptaddress]{revtex4}

\usepackage{graphicx,eucal,amssymb}

\makeatletter
\newcommand{\mylistoffigs}{%begin lof.def.
  \section*{\hspace{-\parindent}%
  FIGURE CAPTIONS}\@starttoc{fig}
}%endof lof.def.
\newcommand{\figcap}[1]{%begin fig.cap.def.
  \refstepcounter{figure}
  \par\vspace*{1in}\centerline{\textbf{FIG.~\thefigure}} \newpage
        \addcontentsline{fig}{fig}{\noindent\textbf{FIG.~\thefigure:}~#1}
}%endof fig.cap.def.
\newcommand{\l@fig}[2]{\par\noindent#1}
\makeatother

\begin{document}
\ifx\href\undefined\else\hypersetup{linktocpage=true}\fi
\sloppy

\title{Quantum Process Tomography of the Quantum Fourier Transform\medskip}

\author{Yaakov S. Weinstein}
\altaffiliation[Present address: ]{\setlength{\baselineskip}{12pt} Center for Computational Materials Science, Naval Research Laboratory, Washington, DC 20375, USA}
\author{Timothy F. Havel}
\author{Joseph Emerson}
\altaffiliation[Present address: ]{\setlength{\baselineskip}{12pt} Perimeter Institute for Theoretical Physics, Waterloo, Ontario N2J 2W9, CANADA}
\author{Nicolas Boulant}
\affiliation{Massachusetts Institute of Technology, Dept. of Nuclear Engineering, Cambridge, MA 02139, USA}
\author{Marcos Saraceno}
\affiliation{Unidad de Actividad Fisica, Tandar, Comisi\'on Nacional de Energia At\'omica, 1429, Buenos Aires, ARGENTINA}
\author{Seth Lloyd}
\affiliation{d'Arbeloff Laboratory for Information Systems and Technology, Dept. of Mechanical Engineering, Massachusetts Institute of Technology,  Cambridge, MA 02139, USA\vspace{0.5\baselineskip}}
\author{David G. Cory}
\affiliation{Massachusetts Institute of Technology, Dept. of Nuclear Engineering, Cambridge, MA 02139, USA}

\pacs{\vspace{-24pt} 03.67.Lx, 82.56.-b, 83.85.Ns}

\begin{abstract} \medskip \setlength{\baselineskip}{15pt}
The results of quantum process tomography on a three-qubit nuclear 
magnetic resonance quantum information processor are presented,
and  shown to be consistent with a detailed model of the
system-plus-apparatus used for the experiments. 
The quantum operation studied was the quantum Fourier transform, which 
is important in several quantum algorithms and poses a rigorous test 
for the precision of our recently-developed strongly modulating control 
fields. The results were analyzed in an attempt to decompose the 
implementation errors into coherent (overall systematic), incoherent 
(microscopically deterministic), and decoherent (microscopically random) 
components. This analysis yielded a superoperator consisting of a 
unitary part that was strongly correlated with the theoretically expected 
unitary superoperator of the quantum Fourier transform, an overall 
attenuation consistent with decoherence, and a residual portion that 
was not completely positive - although complete positivity is required 
for any quantum operation. By comparison with the results of 
computer simulations, the lack of complete positivity was shown to be 
largely a consequence of the incoherent errors during the quantum 
process tomography procedure. These simulations further showed that
coherent, incoherent, and decoherent errors can often be identified by
their distinctive effects on the spectrum of the overall superoperator.
The gate fidelity of the experimentally determined superoperator was $0.64$,
while the correlation coefficient between experimentally determined
superoperator and the simulated superoperator was $0.79$; most of
the discrepancies with the simulations could be explained by the
cummulative effect of small errors in the single qubit gates.
\end{abstract}

\maketitle

\section{Introduction}

In order to develop larger and more powerful quantum
information processing devices, it is essential to
quantify the precision with which they can be controlled.
This information is generally reported as a single number,
the fidelity of the operation \cite{Schu,Grum,E}.
Although fidelity is a reasonable measure of control,
it gives experimentalists little useful information about what 
went wrong or how to improve their control over the quantum system.
Quantum process tomography (QPT) \cite{C+N,PCZ,DM} provides additional
information that may be useful in this regard, by yielding an
estimate of the quantum operation that was actually implemented.
This in turn provides a stringent check on the completeness and accuracy
of the system-plus-apparatus model used to design the implementation.
It is nevertheless a challenging task to interpret the
deviations of this estimate from the superoperator implied
by the model in terms of specific defects in the model.
Additional errors introduced during the QPT procedure
itself further complicate the analysis of the QPT results.
In this paper we explore these issues via a concrete case study,
in which QPT is performed on a previously reported three-qubit
quantum Fourier transform (QFT) implemented using a nuclear
magnetic resonance (NMR) quantum information processor \cite{YSW}.

The dynamics of an isolated quantum system are described by the 
Schr\"odinger equation, which gives rise to an $N\times N$unitary
operator, where $N$ is the dimension of system's Hilbert space.
Open quantum systems, however, generally interact with an
inaccessible environment and thereby undergo decoherence \cite{C+N}.
Furthermore, in the case of expectation value measurements
as in liquid-state NMR, each element of the statistical
ensemble may undergo a slightly different unitary operation which,
though nonrandom, is difficult to distinguish from decoherence \cite{CP,Marco}.
The statistics of measurements on open quantum systems
are generally described by an $N\times N$ density operator,
and the evolution of a density operator under an incoherent
distribution of Hamiltonians and/or interactions with an environment,
although non-unitary, remains linear and is described by a superoperator.
The goal of quantum process tomography is to determine this superoperator.

Methods for implementing QPT have been presented
Refs.~\cite{C+N,PCZ,DM,Leung,Tim}, and two-qubit NMR implementations
of QPT have previously been reported \cite{Andrew,Nick}.
In the present paper, QPT is carried out with the aim of validating
the mathematical model of the system-plus-apparatus used to design
the NMR implementation of a multi-qubit, entangling unitary operation, 
and to identify the types and strengths of the errors that occurred. 
We have found it useful to classify the errors as coherent, incoherent, 
and decoherent, because each class is related to specific short-comings 
in the experimental implementation. Coherent errors are systematic 
errors in the net unitary operation that was actually implemented.
Incoherent errors refer to unwanted unitary evolution which is not
uniform across the ensemble of spin systems in the NMR sample,
so that even though each member of the ensemble undergoes strictly unitary
evolution, the evolution of the ensemble averages appears non-unitary.
The effects of incoherence are reversible, at least in principle,
and knowledge of the coherent and incoherent errors can be
used to design better quantum gates (unitary operations).
In contrast, decoherent errors are due to unknown interactions
with an inaccessible environment, so they are not reversible
and can be eliminated only by relatively costly changes to the
apparatus or the way in which information is encoded within it.  

An important benefit of QPT is that it poses a rigorous
test of the accuracy of the mathematical model of the
system-plus-apparatus used to design and interpret the experiments.
This is done by comparing the experimental results of QPT to the results of
computer simulations of the complete QPT procedure, based upon this same model.
Simulations based on the model used here were also used in designing
the strongly modulating control fields by which both the desired
unitary operation  as well as all the unitary operations
needed for the QPT procedure were implemented \cite{Marco}.
This was done by minimizing the difference between the desired qubit
rotation operator and the quantum operation obtained by simulating
the effect of the control fields on the spins in the molecule
used for the experiments. As a result, any incorrectness in
the model directly affects the reliability of the experiments,
but in ways that, by definition, differ from the simulations.
This can suggest ways to improve the model, after which further
simulations will pinpoint the remaining experimental errors.

\begin{figure}[H] \newpage
\includegraphics[width=8.5cm]{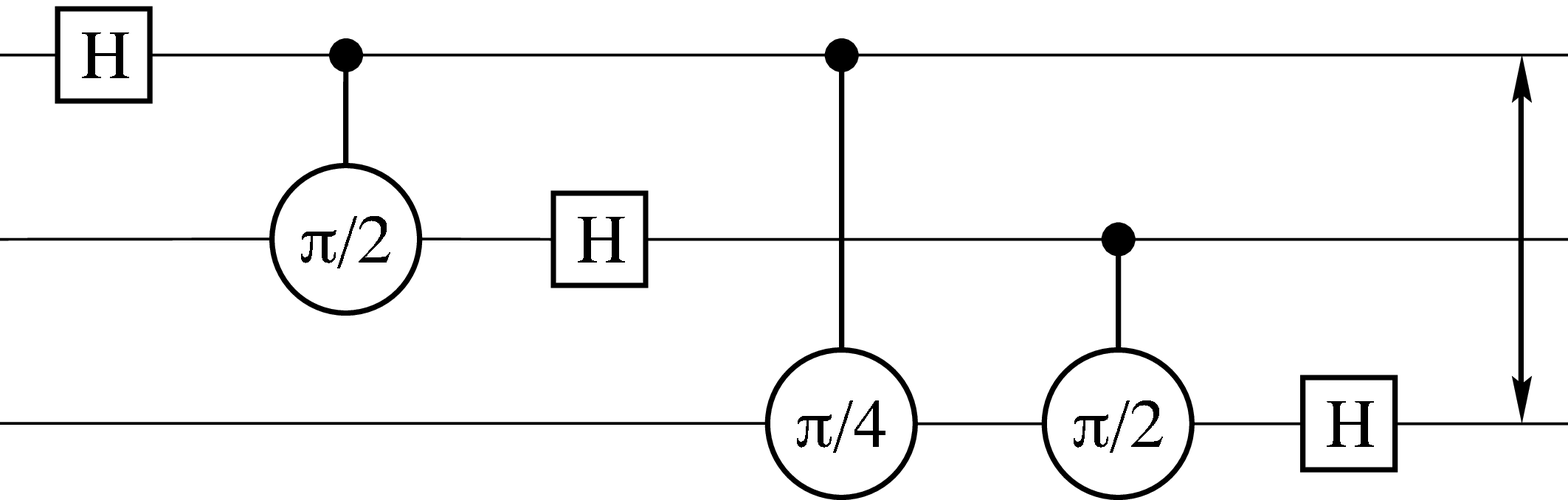}
\figcap{ 
Circuit diagram for implementation of the quantum Fourier transform for 
three qubits. For each qubit $j$, starting with the most significant, a 
series of conditional phase gates are implemented between qubit $j$ and 
all qubits more significant than $j$, followed by a Hadamard ($\mathrm H$)
on $j$. The  amount of phase added is $\theta_{jk} = \pi/2^{j-k}$.
A bit reversal (two-headed arrow) completes the QFT.
} \label{QFTCircuit}
\end{figure}

The operation on which QPT was performed is the quantum Fourier transform (QFT).
The QFT constitutes a key subroutine in several quantum algorithms
\cite{Shor,Josza} as well as in certain methods for simulating 
quantum dynamics on a quantum computer \cite{Zalka,Sch}.  
In algorithms such as Shor's factoring algorithm, the QFT is used to 
extract periodic features of wave functions, while in simulations of
quantum dynamics it is used to move between the position and momentum 
representations. The QFT is defined as follows:
\begin{equation}
U_\mathsf{QFT}|x\rangle = \frac{1}{\sqrt{N}} \sum^{N-1}_{x'=0} e^{2 \pi \mathrm ixx'/N}|x'\rangle.
\end{equation}
The QFT has been expressed  \cite{Cop,Price} as a sequence of one-qubit 
Hadamard gates $H_j$, which rotate the $j$th qubit from a computational basis 
state to an equal superposition of computational basis states, and two-qubit 
conditional phase gates $B_{jk}(\theta)$, which rotate the phase of qubit $k$
by $\theta$ if qubit $j$ is in the state $|1\rangle$. In this notation, the
complete gate sequence of the three-qubit QFT is (reading from right-to-left) 
\begin{eqnarray}
\label{QFTgates}
\begin{array}{c}
Swap_{13}\, H_3\, B_{23}\!\left(\frac{\pi}{2}\right) B_{13}\!\left(\frac{\pi}{4}\right) H_2\, B_{12}\!\left(\frac{\pi}{2}\right) H_1 ~,
\end{array}
\end{eqnarray}
where $Swap_{jk}$ is a swap gate between qubits $j$ and $k$
(see Fig.~\ref{QFTCircuit}). 
This gate sequence has been implemented via NMR to demonstrate the ability of 
the QFT to extract periodicity \cite{YSW}, and as part of the quantum baker's 
map \cite{Baker}.

The remainder of this paper is organized as follows.
Section II gives an overview of the experimental
and computational procedures used for QPT,
together with the metrics by which the results
were compared to those of the theoretical model.
Section II describes incoherent errors and their effects on QPT,
in particular the apparent lack of complete positivity of the results,
while section III describes the experimental system
and procedures used to implement QPT in detail.
Section IV presents a complete description of the
system-plus-apparatus model by which the results of QPT were interpreted.
This is followed in section V by an overview of the experimental results,
and in section VI by a detailed comparison
of the results with the model's predictions.
Finally, section VII contains an analysis of the discrepancies
between the experimental and simulated results,
with the goal of determining their probable origins.
The paper concludes with brief discussion of the implications of our
findings for the field of quantum information processing as a whole.

\section{Quantum Process Tomography} \label{sec:qpt}

\begin{figure}[H] \newpage
\includegraphics[width=8.5cm]{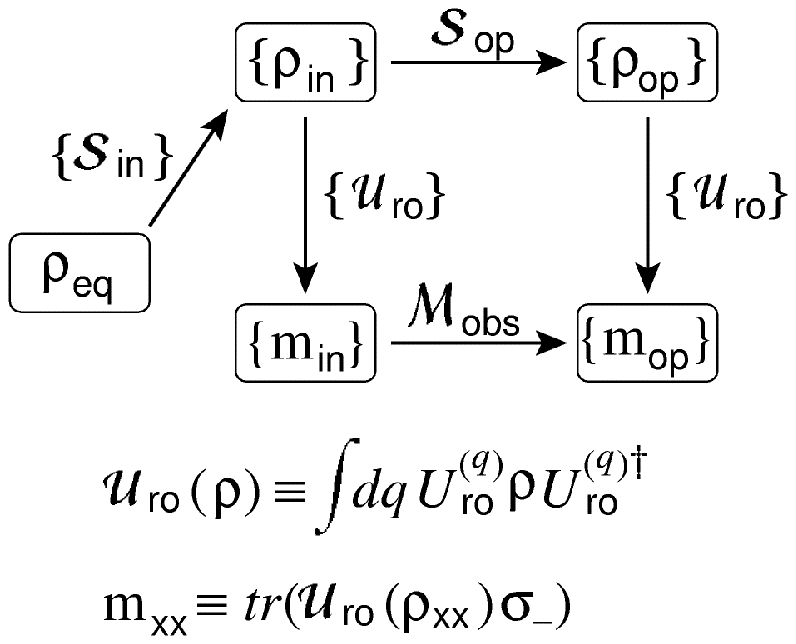}
\figcap{
A schematic of the QPT implementation used in this paper,
together with two equations describing how the readout operations
$\mathcal U_\mathsf{ro}$ and measurements $\mathrm m_\mathsf{xx}$
($\mathsf{xx} = \mathsf{in}, \mathsf{op}$) are obtained from the
ensemble of quantum systems described by the density matrix $\rho$
($\sigma_- \equiv \frac12 ( \sigma_\mathsf x - \mathrm i \sigma_\mathsf y)$).
A set $\{\mathcal S_\mathsf{in}\}$ of (not necessarily unitary) 
operations is performed, each in a separate experiment, on the
equilibrium state $\rho_\mathsf{eq}$ of the spin system to create a 
complete set $\{\rho_\mathsf{in}\}$ of input states. Each input state is then 
determined by quantum state tomography, i.e.~by repetition of the experiment 
with different readout pulses, $\{\mathcal U_\mathsf{ro}\}$, appended to each 
repetition. The readout pulses rotate unobservable components of the
density matrix into observable components, the mean values
$\{{\rm m}_\mathsf{in}\}$ of which are equal to the unobservables'
mean values before rotation. Measurement of these mean values thus 
allows for the reconstruction of the input state that was actually created.
Next, the set of input states is recreated one state at a time, and the operation
$\mathcal S_\mathsf{op}$ of interest (in this paper the QFT) applied to each in turn.
This gives a complete set of output states, $\{\rho_\mathsf{op}\}$.
Once again, readout of each  $\rho_\mathsf{op}$ requires that
the experiment be repeated followed by different readout operations
$\{\mathcal U_\mathsf{ro}\}$. The mean values of the complete
set of observables $\{{\rm m}_\mathsf{op}\}$ that is measured
allows reconstruction of the set of output states actually created.
Finally, these estimates of $\{\rho_\mathsf{op}\}$, together with
the earlier estimates of $\{\rho_\mathsf{in}\}$, are used to reconstruct 
an estimate of the operation $\mathcal S_\mathsf{op}$ via Eqn.~(\ref{Sexp}).
This reconstructed operation is not expected to be unitary due to decoherence 
during the application of $\mathcal S_\mathsf{op}$ as well as errors in the 
state tomography procedure. Also, due to incoherent errors across the 
ensemble involved in NMR experiments, the supermatrix $\mathcal M_\mathsf{obs}$
estimated via this QPT procedure may not correspond to a completely positive,
trace preserving superoperator.
} \label{QPT}
\end{figure}

There are several methods of performing QPT outlined in the literature.
Some of these methods \cite{DM,Leung} require increasing the Hilbert space 
size beyond that of the system whose dynamics are to be studied. This is 
unappealing for current experimental studies of quantum information 
processing where qubits are at a premium. The procedure used in this work 
(see Fig.~\ref{QPT}) is similar to those given in Refs.~\cite{C+N,PCZ}, 
and results in an $N^2 \times N^2$ complex-valued matrix, hereafter referred 
to as a ``supermatrix.'' Because of the inevitable experimental errors made 
in the QPT procedure, this supermatrix will not, in general, correspond to 
any completely positive and trace-preserving superoperator (ergo quantum 
operation), unitary or otherwise. Therefore, we use methods recently 
introduced by Havel \cite{Tim,Nick} to obtain the best least-squares fit 
to the measured supermatrix that does correspond to such a quantum operation. 
This is expected to yield a better estimate of the superoperator that
was actually implemented, since it will subsequently be shown
that the non-completely positive part of the superoperator
is largely due to errors committed during the readout steps.

The QPT procedure in Fig.~\ref{QPT} requires implementation of the desired 
unitary operation on a complete set of $N^2$ known input states, followed by 
determination of all of the resulting output states via quantum \emph{state}
tomography. Once all of the experimental input and output states have 
been completely determined, and given that the input states have been
chosen to be linearly independent, the experimental supermatrix can be 
calculated as 
\begin{equation}
{\mathcal M}_\mathsf{obs} ~=~ {\mathcal R}_\mathsf{out}\,
{\mathcal R}_\mathsf{in}^{-1} \label{Sexp} ~,
\end{equation}
where ${\mathcal R}_\mathsf{in}$ and ${\mathcal R}_\mathsf{out}$
are supermatrices whose columns are the ``columnized'' experimental
input and output density matrices $col(\rho_\mathsf{in})$ and
$col(\rho_\mathsf{op})$, respectively, as determined by state tomography (see below).
In quantum information processing, the Hilbert space is usually a tensor
product of the $2$-dimensional Hilbert spaces of its constituent qubits,
so that $N = 2^n$ grows exponentially with the number of qubits $n$.
As a result, QPT can in practice be carried out on only a few qubits
at a time. Nevertheless, even a large quantum computer is expected
to be based upon such local operations, so a complete analysis
of small implementations is a prerequisite to understanding
the issues involved in the control of larger systems.

We now introduce the measures of state and gate
fidelities that were used to summarize the results of QPT.
The accuracy with which the initial states were created was
quantitated by the correlation between the desired input state
and the one determined by state tomography \cite{E}, i.e.
\begin{equation}
\label{proj}
C(\rho_\mathsf{th},\rho_\mathsf{in}) ~=~ \frac{\mathrm{tr}(\rho_\mathsf{th}\rho_\mathsf{in})}
{\sqrt{\mathrm{tr}(\rho_\mathsf{th}^2) \mathrm{tr}(\rho_\mathsf{in}^2)}} ~,
\end{equation}
where $\rho_\mathsf{th}$ and $\rho_\mathsf{in}$ are the traceless parts of the
density matrices of the desired and measured input states, respectively.
The input states were prepared from the equilibrium spin state
by means of suitable non-unitary operations, $\mathcal S_\mathsf{in}$,
and the magnitudes of their traceless parts taken as the
reference against which all subsequent losses of coherence
(or magnetization in NMR) due to non-unitary evolution were measured.
This was done by scaling the correlation of the output states by a factor which
measures the loss of coherence \cite{E}, yielding the so-called attenuated correlation
\begin{equation}
C_\mathsf A(\rho_\mathsf{th}, \rho_\mathsf{op}) ~=~ C(\rho_\mathsf{th}, \rho_\mathsf{op})\sqrt{\frac{\mathrm{tr}(\rho_\mathsf{op}^2)}{\mathrm{tr}(\rho_\mathsf{in}^2)}} ~,
\end{equation}
where $\rho_\mathsf{th}$ and $\rho_\mathsf{op}$ are the traceless parts of the density
matrices of the theoretical and measured output states, respectively,
and $\rho_\mathsf{in}$ that of the corresponding input state from above.

The average state correlation and average attenuated state
correlation are reasonable measures of the overall fidelity
of the implemented quantum operation, but a measure that is
more clearly independent of the choice of initial states
is the correlation between the superoperator matrices, i.e. 
\begin{equation}
C({\mathcal S}_\mathsf{th},\, {\mathcal S}_\mathsf{op}) ~\equiv~
\mathrm{tr}({\mathcal S}_\mathsf{th}^{\dag}{\mathcal S}_\mathsf{op}) \text{\Large$/$}
\! \sqrt{\mathrm{tr}({\mathcal S}_\mathsf{th}^\dag{\mathcal S}_\mathsf{th})
\, \mathrm{tr}({\mathcal S}_\mathsf{op}^\dag{\mathcal S}_\mathsf{op})} ~.
\end{equation}
Since this quantity is not sensitive to the overall loss of magnetization,
we will also use the attentuated correlation between supermatrices, namely
\begin{equation}
C_A({\mathcal S}_\mathsf{th},{\mathcal S}_\mathsf{op}) ~\equiv~
\mathrm{tr}({\mathcal S}_\mathsf{th}^{\dag}{\mathcal S}_\mathsf{op})
\text{\large$/$} \mathrm{tr}({\mathcal S}_\mathsf{th}^{\dag}{\mathcal S}_\mathsf{th})
~=~ F_\mathsf e({\mathcal S}_\mathsf{th}^{\dag}{\mathcal S}_\mathsf{op}) ~,
\label{eqn:fid}
\end{equation}
where $F_\mathsf e$ is the entanglement fidelity defined by Schumacher \cite{Schu}.
Accordingly, we will refer to this quantity as the gate fidelity.
It can be shown that this fidelity satisfies
$\bar C_A(\rho_\mathsf{th},\rho_\mathsf{op}) \ge
C_A({\mathcal S}_\mathsf{th},{\mathcal S}_\mathsf{op})$,
with equality if $\mathrm{tr}(\rho_\mathsf{in}^2) = \mathrm{tr}(\rho_\mathsf{th}^2)$
for all the input states \cite{E}.

Our analysis utilizes two different representations of quantum operations, 
the supermatrix representation and the Kraus operator sum representation. 
The supermatrix $\mathcal S$ operates on the density matrix as
\begin{equation}
\mathit{col}(\rho_\mathsf{fin}) ~=~ {\mathcal S}\;\mathit{col}(\rho_\mathsf{ini}) ~,
\end{equation}
where $\rho_\mathsf{ini}$ and $\rho_\mathsf{fin}$ are the initial 
and final density matrices, and the ``$col$'' operation
stacks the columns of the density matrices on top of 
one another in left to right order. This results in an 
$N\times N$ matrix becoming a column vector of length $N^2$. 
We will primarily work in two different supermatrix bases, the computational 
(or Zeeman) basis and the product operator basis. In the computational 
basis the rows and columns of the $2^n \times 2^n$ density matrix are 
labeled by the binary expansion of their indices from $|0\dots 0\rangle$ to 
$|1\dots 1\rangle$. The product operator basis is related to the 
structure of the NMR Hamiltonian and rows and columns of the supermatrix 
in this basis are labeled as 
$I_2\otimes\dots\otimes I_2$, $I_2\otimes\dots\otimes I_2\otimes\sigma_\mathsf{x}$,
to  $\sigma_\mathsf{z}\otimes\dots\otimes \sigma_\mathsf{z}$, where ``$\otimes$''
denotes tensor multiplication, the $\sigma$'s are the standard Pauli matrices,
$I_2$ is the $2 \times 2$ identity matrix and each term has $n$ factors
(see Ref.~\cite{Tim2} for details).

The superoperator of a quantum operation can also be
expressed as a Kraus operator sum. This sum is comprised 
of a set of $N \times N$ matrices, $A_k\,$, such that
\begin{equation}
\rho_\mathsf{fin} ~=~ {\sum}_k\, A_k\, \rho_\mathsf{ini\,} A_k^{\dag}
\end{equation}
where the sum may require as many as $N^2$ terms.
It is easily seen that a Kraus operator sum 
preserves the Hermiticity of $\rho_\mathsf{ini}$, and
that it preserves the trace if and only if
\begin{equation}
{\sum}_k\, A_k^{\dag}\, A_k ~=~ 1 ~.
\end{equation}
If the operation is purely unitary there is only one
Kraus operator, which is just the unitary operator in question.
In general, however, there are an infinite number of equivalent Kraus operator 
representations of a given quantum operation. The least-squares fitting 
procedure described below yields a Kraus operator sum representation in 
which the Kraus operators are both mutually orthogonal, minimum in number
and sorted by the size of the contribution each makes to $\rho_\mathsf{fin}$.

\section{Incoherent Errors} \label{sec:incoh}

A superoperator is completely positive
if it admits a Kraus operator sum representation.
Although this condition holds for the physical processes usually studied 
in the quantum mechanics of open systems, including unitary operations and 
decoherence via weak interactions with a Markovian environment, it is not 
difficult to construct situations in which not only complete positivity, 
but even the distinction between a state and its transformations breaks down 
\cite{P,S1,S2,N}. Such situations can only arise when the initial state of
the system is not pure, but can be represented by a probability distribution
over an ensemble of pure states with density matrix $\rho_\mathsf{ini}$.
Suppose the ensemble's probability distribution depends on some 
classical parameter $c$ (usually space or time) and that $\rho_\mathsf{ini}$ also 
represents an average over $c$ (see below for a concrete example). Then if 
the applied transformation $\mathcal K$ likewise depends on $c$, so that 
the transformation is \emph{correlated} with the states in the ensemble, 
then the final density matrix $\rho_\mathsf{fin}$ will generally not be equal to 
the result of applying the average of $\mathcal K$ over $c$ to $\rho_\mathsf{ini}$. 
Indeed there will usually be no superoperator which maps every possible 
$\rho_\mathsf{ini}$ to the correct $\rho_\mathsf{fin}$.
In principle one could define a non-linear transformation that
produces this mapping, but this is not very useful in practice because
the large amount of information needed to define a general non-linear
transformation is neither readily available nor easy to work with.

Incoherent errors are precisely those which depend upon some classical 
parameter that labels the members of an ensemble, and so generate a 
correlation between the states in the ensemble and this parameter.
In the NMR experiments described below, the main source
of incoherent errors is the spatial inhomogeneity in
the RF (radio-frequency) field over the sample volume.
This may be expressed by writing the Hamiltonian for the
interaction of the spins with the RF field in the form
\begin{equation}
H_\mathsf{RF}(t; \vec r\,) ~=~ \alpha(t)\, \gamma B_1(\vec r\,)\,
e^{-{\rm i} \Delta\omega\, t\sigma_\mathsf{z}/2}\, \sigma_\mathsf{x}\,
e^{\,{\rm i} \Delta\omega\, t\sigma_\mathsf{z}/2} ~,
\label{eq:Hrf}
\end{equation}
where $\alpha(t)$ is the envelope of the field, $\gamma$ is the
gyromagnetic ratio of the spins, $B_1(\vec r\,)$ is the magnitude
of the field at the point $\vec r\,$, $\Delta \omega$ is
the difference between the frequency of the RF field and the
resonance frequency of the spins in the static magnetic field $B_0$,
and the $\sigma$'s are the usual Pauli matrices.
Thus, the unitary operation $U = \exp\big( \int {\rm d}t\, H_\mathsf{RF} \big)$
depends upon the position $\vec r\,$ in the sample as well as time,
i.e.~$U = U(t;\vec r\,)$, as does the resulting spin state
$\rho(t;\vec r\,) = U(t;\vec r\,) \rho(0;\vec r\,) U^\dag(t;\vec r\,)$.
The equilibrium state of the spins in the static field $B_0$
is independent of $\vec r\,$ over the sample volume, so
we may take $\rho(0;\vec r\,) = \rho(0)$ to be constant.
The density operator that is observed is
the integral over the sample volume $V$, i.e.
\begin{equation}
\rho(t) = \int_V \mathrm d \vec r\; \rho(t;\vec r\,) = \int_V \mathrm d \vec r\; U(t;\vec r\,) \rho(0) U^\dag(t;\vec r\,) \;\equiv\; {\mathcal S}(t) \rho(0) .
\end{equation}
where $\mathcal S(t)$ denotes the net superoperator of the actual quantum
operation implemented. We can express this in matrix form as \cite{Tim}
\begin{equation}
\mathit{col}\big( {\mathcal S}(t) \rho(0) \big) = \bigg( \int_V \mathrm d \vec r\;
\overline U(t;\vec r\,) \otimes U(t;\vec r\,) \bigg) \mathit{col}\big( \rho(0) \big) ,
\end{equation}
where $\overline U$ is the complex conjugate of $U$ and
the ``$col$'' operator maps $N\times N$ matrices to
$(N^2)$-dimensional column vectors, as described above.

\begin{figure}[H] \eject
\includegraphics[width=8.5cm]{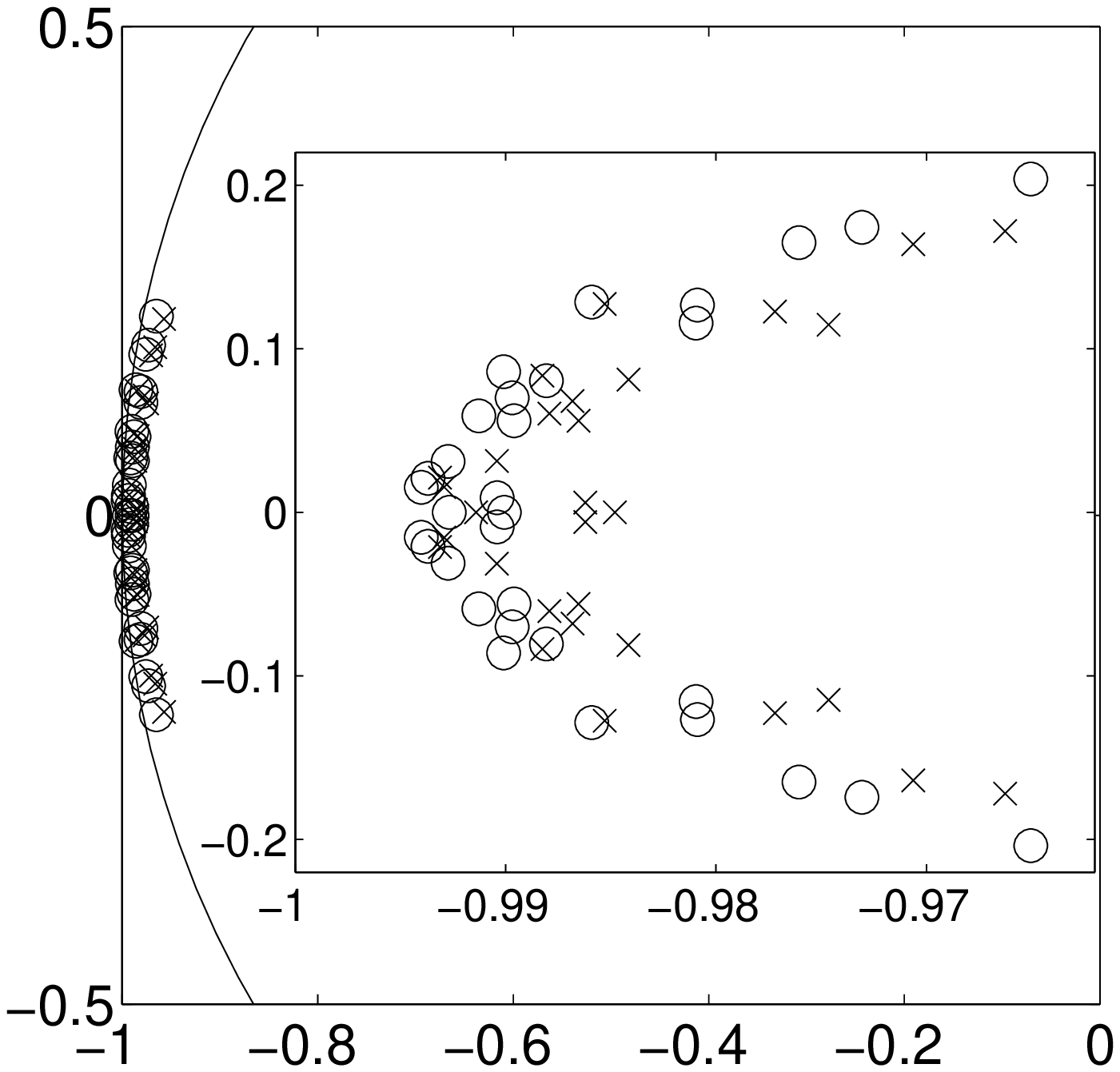}
\figcap{Eigenvalues of simulated supermatrices in the 
complex plane. The main figure shows the location of the eigenvalues with 
respect to the unit circle, while the subplot shows the eigenvalue locations 
in greater detail. This simulation was done for alanine (see text), and 
includes both coherent errors due to imperfect pulses and incoherent errors 
due to the inhomogeneous RF power across the sample. The eigenvalues indicated
by circles (\protect\raisebox{2pt}{\tiny$\bigcirc$}) are of the product
$\mathcal S_{180\mathsf x}^{1,2}\,{\mathcal S}_{90\bar{\mathsf y}}^{1}$ of
the supermatrix describing an evolution ${\mathcal S}_{90\bar{\mathsf y}}^{1}$
of a $90^\circ$ pulse on spin 1 about the negative $\mathsf y$-axis
times a second supermatrix $\mathcal S_{180\mathsf x}^{1,2}$ for
a $180^\circ$ rotation of spins 1 and 2 about the $\mathsf x$-axis. 
The eigenvalues indicated by crosses ($\times$) are for the supermatrix
${\mathcal S}_\mathsf{both}$ describing the net evolution after
concatenating these pulses together in the simulator.
Due to the presence of incoherence the eigenvalues are not the same.
In particular, the presence of incoherence leads to an apparent decoherence,
as evidenced by a reduction in the eigenvalue magnitudes.
The average eigenvalue reduction is 1.2\% for ${\mathcal S}_\mathsf{both}$ and 
0.9\% for $\mathcal S_{180\mathsf x}^{1,2}\, {\mathcal S}_{90\bar{\mathsf y}}^{1}$.
} \label{TwoPulse} 
\end{figure}

Describing the evolution of an ensemble by integrating over
its spatial degrees of freedom works well for a single operation.
When a second operation $\mathcal S_2$ is applied following the first 
$\mathcal S_1$, however, we must take the spatial correlations produced 
by the first into account in computing the correct overall superoperator.
In terms of the foregoing equations, this may be expressed as
\begin{eqnarray}
{\mathcal S}_2\, {\mathcal S}_1 &=&
\int_V \mathrm d \vec r_2\; \Big( \overline U_2(\vec r_2) \otimes U_2(\vec r_2) \Big)
\int_V \mathrm d \vec r_1\; \Big( \overline U_1(\vec r_1) \otimes U_1(\vec r_1) \Big)
\nonumber \\
&=& \int_V \mathrm d \vec r_2\int_V \mathrm d \vec r_1\;
\Big( \overline U_2(\vec r_2) \overline U_1(\vec r_1) \Big)
\otimes \Big( U_2(\vec r_2) U_1(\vec r_1) \Big)
\\ \nonumber
&\ne& \int_V \mathrm d \vec r\; \Big( \overline U_2(\vec r\,) \overline U_1(\vec r\,) \Big)
\otimes \Big( U_2(\vec r\,) U_1(\vec r\,) \Big) ~=~ {\mathcal S} ~,
\end{eqnarray}
where $\mathcal S$ is the true superoperator of the combined operations.
Clearly, a similar result could be obtained if we were integrating over time,
although that is not relevant to the experiments described in this paper.
A simulation of an example of the lack of composibility of superoperators
describing incoherent errors is shown in Fig.~\ref{TwoPulse}.

It follows that the $N^2 \times N^2$ supermatrix produced by our QPT 
procedure (Fig.~\ref{QPT}) cannot be expected to precisely correspond to 
any physical process, and it will usually even fail to define a completely 
positive and trace-preserving superoperator. In this event, an improved 
estimate of the supermatrix of the quantum process that is realized by 
the QFT implementation can be obtained by making the smallest possible 
change to the supermatrix derived from QPT so as to make it completely 
positive and trace-preserving (CPTP).

One can, of course, always convert a Kraus operator sum into an
equivalent supermatrix via the ``$col$'' operation introduced above,
\begin{equation}
\mathit{col}\Big( {\sum}_k A_k\, \rho_\mathsf{ini}\, A_k^{\dag} \Big) ~=~
\Big( {\sum}_k \overline A_k \otimes A_k \Big) \mathit{col}\big( \rho_\mathsf{ini} \big) .
\end{equation}
To go the other direction, i.e.~to take a supermatrix $\mathcal S$ 
and convert it into an equivalent Kraus operator sum, we first rearrange
its elements into a Hermitian supermatrix called the Choi supermatrix \cite{Tim},
\begin{equation}
{\mathcal T} ~=~ \sum_{ij = 0}^{N-1} \big( E_{ij}\otimes I_N){\mathcal S}(I_N\otimes E_{ij} \big),
\label{Choi}
\end{equation}
where $I_N$ is the $N\times N$ identity matrix and $E_{ij}$ is
a matrix of zeros except for a $1$ in the $(i,j)$th position.
Then, if ${\mathcal T} = \sum_k \lambda_k v_k v_k^\dag$ is the spectral 
decomposition of ${\mathcal T}$ and we assume that all the eigenvalues 
$\lambda_k \ge0$, our Kraus operators $A_k$ may be shown to be
\begin{equation}
\mathit{col}\big( A_k \big) ~=~ \sqrt{\lambda_k}\, v_k \qquad(k = 1,\ldots,N) .
\end{equation}

It follows from the above that a superoperator $\mathcal S$ is completely
positive if and only if its Choi supermatrix is positive semi-definite.
In the event that an experimentally determined supermatrix $\mathcal S_\mathsf{obs}$
does not have a positive semi-definite Choi supermatrix, it has been shown 
\cite{Tim} that the completely positive superoperator $\mathcal S'$ closest to
it in the least-squares sense may be obtained simply by setting the negative
eigenvalues of its Choi supermatrix to zero to get a new matrix ${\mathcal T}'$,
and mapping it back to $\mathcal S'$ via Eqn.~(\ref{Choi}) with $\mathcal S$ and 
${\mathcal T}$ swapped. This procedure, however, will result in a 
supermatrix $\mathcal S'$ that does not preserve the trace of the 
density matrix. That condition can be reimposed by subtracting
\begin{equation}
\Delta {\mathcal T}' ~\equiv~ N^{-1} col(I_N)\,
col^\dag(I_N)\, {\mathcal T}' - I_{N\times N}
\end{equation}
from ${\mathcal T}'$, with the result that ${\mathcal T}'$ is no longer 
positive semi-definite. It has further been shown, however, that iterating on 
these two procedures generates a sequence of supermatrices which converges to 
the Choi supermatrix corresponding to the superoperator that is closest to
$\mathcal S$ and is \emph{both} trace preserving and completely positive.
This is described for the experimental results in the Section \ref{sec:disc}.

In order to quantify the extent to which a supermatrix violates the completely
positive requirement, we define the positivity as the ratio of the sum 
of eigenvalues of the Choi matrix corresponding to the supermatrix 
over the sum of positive eigenvalues of the said Choi matrix.
For a completely positive superoperator the positivity is equal to 1,
while the presence of negative Choi eigenvalues causes the positivity 
to be less than 1.

\section{The Spin System and the Experiments} \label{sec:exp}
The experiments were implemented on a three-qubit NMR quantum information 
processor \cite{Cory1,Jones}. The three qubits used were the three carbon
spins in molecules of $^{13}$C-enriched alanine in an aqueous solution.
The internal Hamiltonian of this system has the form
%\begin{equation}
%H_\mathsf s ~=~ \pi\nu_1\, \sigma_\mathsf{z}^1+\pi\nu_2\, %\sigma_\mathsf{z}^2+\pi\nu_3\, \sigma_\mathsf{z}^3+ 
%\frac{\pi}{2}\big( J_{1,2}\, \vec{\sigma}^1\cdot\vec{\sigma}^2+ J_{1,3}\,
%\vec{\sigma}^1\cdot\vec{\sigma}^3+ J_{2,3}\, \vec{\sigma}^2\cdot\vec{\sigma}^3
%\big) ,
%\end{equation}
\begin{equation} \label{H_small_s}
H_\mathsf s ~=~ \pi \sum_{i=1}^3\, \nu_\mathsf C^i\: \sigma_{\mathsf C,\mathsf z}^i \:+\: \frac\pi2\! \sum_{j>i=1}^3 J_{\mathsf C, \mathsf C}^{\,i,j}\: \vec\sigma_\mathsf C^{\,i} \cdot \vec\sigma_\mathsf C^{\,j} ~,
\end{equation}
where $\nu_\mathsf C^i$ are the Larmor frequencies of the spins
and $J_{\mathsf C,\mathsf C}^{\,i,j}$ are the strengths of the
couplings between them, both in frequency units.
In our indexing scheme, a superscript $1$ labels the carbonyl
carbon of alanine, $2$ the alpha carbon and $3$ the methyl carbon.
A separate time-dependent external Hamiltonian, shown in Eqn.~(\ref{eq:Hrf}),
must be added whenever an RF pulse is applied to rotate qubits. 
In the 7 Tesla magnet used for the experiments, the resonant frequency of 
carbon-13 is approximately 75.468 MHz. Frequency changes, also known as 
chemical shifts, among the spins introduce differences
$\nu_\mathsf C^2 - \nu_\mathsf C^1 = 9456.5$ Hz, 
$\nu_\mathsf C^3 - \nu_\mathsf C^2 = 2594.3$ Hz and
$\nu_\mathsf C^3 - \nu_\mathsf C^1 =  12050.8$ Hz. 
The coupling constants between the three spins are
$J_{\mathsf C,\mathsf C}^{1,2}$ = 54.2 Hz, 
$J_{\mathsf C,\mathsf C}^{2,3}$ = 35.1 Hz, and
$J_{\mathsf C,\mathsf C}^{1,3}$ = $-$1.2 Hz.
In the absence of RF, the $T_1$ relaxation times
of the three spins are all longer than 1.5 s,
while the $T_2$ relaxation times are longer than 400 ms
(see Table \ref{tab:rel} for exact numbers).

To rotate qubits we used the strongly modulating pulses introduced in 
Refs.~\cite{E,Marco}. The first generation of strongly modulating pulses
were designed to perform the desired propagator while refocussing the
spins' known internal Hamiltonian \cite{E}; the second generation 
was designed to also compensate for RF field inhomogeneity \cite{Marco}. 
This was done by including the RF inhomogeneity profile in the simulations
and, hence, in the target function minimized to design the compensated pulses.
Pulses designed in this way performed near-optimally over
the range of RF powers experienced by the ensemble of spins.
While the simulated peak fidelities were lower than with
the uncompensated pulses, the actual performance of the
compensated pulses in the spectrometer was greatly enhanced. 

The method of implementing the QFT via NMR is the same as in our
previous QFT implementations \cite{YSW,Baker}, with the added benefit 
of the RF pulses that compensated for RF-inhomogeneity. The pulse 
sequences for the Hadamard and conditional phase gates are derived 
from an idempotent or correlation operator description of the propagators 
\cite{Shy}. The Hadamard gate pulse sequence is:
\begin{equation}
H_j = \left(\rule{0pt}{3ex}\frac{\pi}{2}\right)_\mathsf y^j
- \left(\rule{0pt}{3ex}\pi\right)_\mathsf x^j.
\end{equation}   
This pulse program reads: apply a pulse that rotates spin $j$ by $90^\circ$ 
about the $\textsf y$-axis, followed by a pulse that rotates $j$ by $180^\circ$ 
about the $\textsf x$-axis. The $B_{jk}$ gate, can be implemented using the 
coupling between qubits and the following pulse sequence:
\begin{eqnarray}
\left(\rule{0pt}{3ex}\pi\right)_{\phi}^{j} - \left(\frac{\theta_{jk}}{2\pi J_{jk}}\right) - \left(\rule{0pt}{3ex}\pi\right)_{\phi}^{j} -
\left(\rule{0pt}{3ex}\frac{\pi}{2}\right)_\mathsf y^{j,k} -
\left(\rule{0pt}{3ex} \frac{\theta}{2}\right)_\mathsf x^{j,k}
- \left(\rule{0pt}{3ex} \frac{\pi}{2}\right)_{\bar{\mathsf y}}^{j,k}.  
\end{eqnarray}
where $\phi$ is an arbitrary phase. The notation 
$\Big( \frac{\theta_{jk}}{2\pi J_{jk}} \Big)$ 
represents a time interval during which $\sigma^j_\mathsf z\sigma^k_\mathsf z$ 
evolution occurs while chemical shifts and all other qubit 
couplings are refocused, while the superscript $j,k$ denotes a pulse which 
rotates only spins $j$ and $k$. The final three pulses in the above sequence 
perform a rotation around the $\textsf z$-axis. In our previous work \cite{YSW} the 
bit reversal was implemented experimentally. Here the bit reversal is achieved 
by simply renaming the bits.

We stress that the various building blocks of the QFT implementation, i.e.~the
individual pulses and evolution periods, remain unchanged for all experiments.
After creation of each initial state a fixed sequence
is applied that is independent of the initial state. 

\begin{figure}[H] \newpage
\vspace{-0.5in}
\includegraphics[height=7.5in]{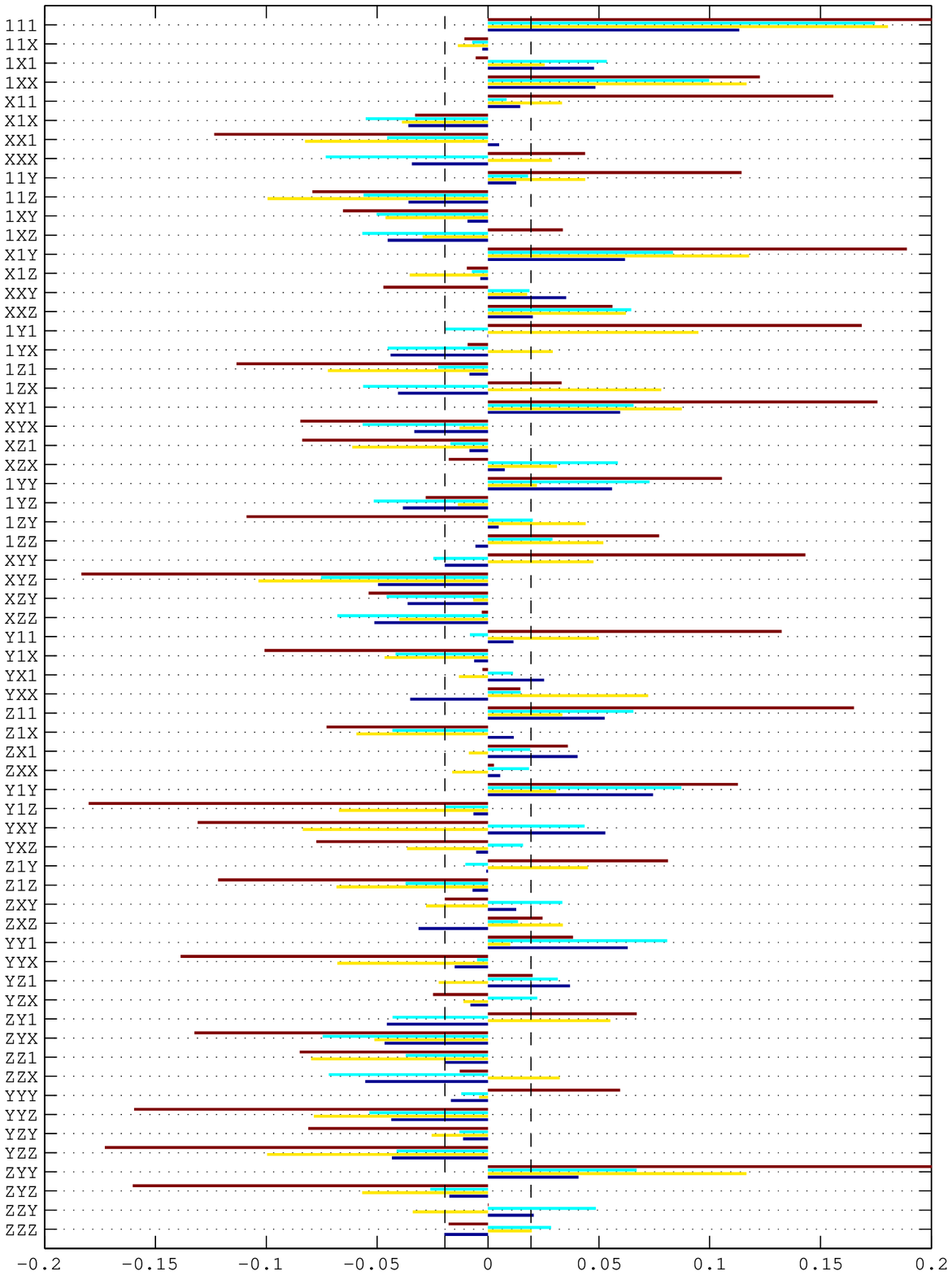}
\figcap{
The differences between the (attenuated) correlations of the output
states and their mean values are displayed for each input state,
which are indicated by the symbols on the left vertical axis with
e.g.~$\mathtt{1XY} \leftrightarrow \sigma_\mathsf x^2 \sigma_\mathsf y^3$.
The simulated-vs-theoretical state correlations are in blue, and
the experimental-vs-theoretical state correlations are in yellow,
while the corresponding attenuated correlations are in cyan and brown,
respectively. The corresponding mean values of the (attenuated)
correlations were $0.89$, $0.82$, $0.83$ and $0.63$, respectively.
The pair of dashed vertical lines about the center mark
one standard deviation for the input state correlations
(not shown), which are expected to be determined primarily
by measurement errors so that any values within these
bounds are certainly almost entirely due to noise.
The larger deviations in the correlations are due
to the propagation of coherent error (see text),
while the yet larger deviations in the attenuated
correlations include the effects of decoherence
(modeled as a uniform attenuation in the simulations).
} \label{AllCorBarH}
\end{figure}

The input states, $\{\rho_\mathsf{in}\}$, used were the 64 product
operator states (neglecting the large but undetectable identity component).
These are $0$ (actually the identity $I_8$), and states such as
$\sigma_\mathsf{x}^1$, $\sigma_\mathsf{z}^3$, and
$\sigma_\mathsf{y}^2\sigma_\mathsf{x}^3$ where the
superscripts represent the spin indices, as above. 
These states were chosen because they are orthogonal and easy to create on a
liquid-state NMR system. We order the states lexicographically starting with
$0$, followed by  $\sigma_\mathsf{x}^3 = I_2\otimes I_2\otimes \sigma_\mathsf{x}$, 
$\sigma_\mathsf{x}^2 = I_2\otimes \sigma_\mathsf{x}\otimes I_2$, and so
on until $\sigma_\mathsf{z}^1\sigma_\mathsf{z}^2\sigma_\mathsf{z}^3 =
\sigma_\mathsf{z}\otimes\sigma_\mathsf{z}\otimes \sigma_\mathsf{z}$
(the complete ordering may be found in Fig.~\ref{AllCorBarH}).
The QFT gate sequence described above is applied to each of these 64 states
to obtain the corresponding set of output states $\{\rho_\mathsf{op}\}$.

In any given three-spin spectrum only 24 of the 64 product operators are 
observable. Thus, to completely reconstruct the state of the system, each 
experiment is repeated 7 times with different readout pulses appended
to the experiment. The real amplitudes of the peaks in the spectrum were then
used as the coefficients of the product operators which were transformed
into the corresponding observable product operators by the readout pulses.
This procedure, known as state tomography \cite{E},
was done for all input and output states.

\section{The System-Plus-Apparatus Model} \label{sec:model}
Quantum process tomography serves experimentalists in two ways.
First, it provides a thorough test of the model of the 
system-plus-apparatus used to design and interpret the experiments.
Second, it provides a complete picture of the net effects of all the errors
made in implementing the experiment and, in conjunction with  the model,
allows these effects to be unravelled into specific defects in the apparatus 
-- and perhaps also the model. Before attempting this latter task, therefore, 
we wish to fully present, discuss, and justify the model we used for QPT 
on the QFT. This model includes the following attributes:
\begin{itemize}
\item The Hilbert space of the system in question and its Hamiltonian.
\item The Hilbert space of the larger system with which the given system interacts coherently, and the Hamiltonian describing these interactions.
\item The relaxation superoperator of the system, and some knowledge of how the larger system relaxes.
\item Bounds on the precision of the classical control fields applied.
\item The distribution of incoherent variations in these fields across the sample.
\end{itemize}
Both the fields and the positions of the spins can be treated classically.
Since the larger system contains only a total of eight spins, we can simulate 
it exactly. Further information, for example the correlation times or full
spectral densities of the noise generators driving relaxation \cite{Ernst},
could have  been included in the model, but the above proved adequate
to explain most of the experimental observations for the QFT.

The first attribute is the Hamiltonian for the three carbon-13
spins of alanine used as qubits for the QFT, which has the form
of $H_\mathsf s$ given in Eqn.~(\ref{H_small_s}).
The larger system includes, in addition to the carbons, the four 
(nonexchanging) hydrogens and the spin-1 nitrogen-14 nucleus in alanine.
The larger system's Hamiltonian has the form
\begin{eqnarray}
H_\mathsf S &=& H_\mathsf s \:+\: \pi \sum_{i=1}^4 \nu_\mathsf H^i\, \sigma_{\mathsf H,\mathsf z}^i \:+\: \pi\, \nu_\mathsf N\, \sigma_{\mathsf N,\mathsf z} \:+\: \frac\pi2\! \sum_{j>i=1}^4 J_{\mathsf H, \mathsf H}^{\,i,j}\: \vec\sigma_\mathsf H^{\,i} \cdot \vec\sigma_\mathsf H^{\,j} ~+ \\ \nonumber
&& \frac\pi2 \sum_{i=1}^4 \sum_{j=1}^3 J_{\mathsf C, \mathsf H}^{\,i,j}\, \sigma_{\mathsf H,\mathsf z}^i \sigma_{\mathsf C,\mathsf z}^j \:+\: \frac\pi2 \sum_{i=1}^3 J_{\mathsf N, \mathsf C}^{\,i}\, \sigma_{\rule{0pt}{1.4ex}
\mathsf N,\mathsf z} \sigma_{\mathsf C,\mathsf z}^{\,i} \:+\:
\frac\pi2 \sum_{j=1}^4\, J_{\mathsf N, \mathsf H}^{\,j}\,
\sigma_{\rule{0pt}{1.4ex}\mathsf N,\mathsf z} \sigma_{\mathsf H,\mathsf z}^{\,j} ~,
\end{eqnarray}
where we have used the well-known fact that couplings between
spins with distinct gyromagnetic ratios can be truncated to just
the secular (i.e.~$\sigma_\mathsf{z} \sigma_\mathsf{z}$) part.

The spin-1 nitrogen has an electric quadrupole moment and hence a short $T_1$,
so its coupling to the  other spins cannot be observed. As a result, the nitrogen
can be omitted from the Hamiltonian, although it still plays a role in the relaxation
of the other  spins. The spin-lattice ($T_1$) relaxation time of the hydrogen atoms, 
on the other hand, is longer than the experiment, so here we must include the 
additional frequency shifts that depend upon their spin states.
In other words, we take the $\sigma_{\mathsf H,\mathsf z}$ to be constants of the 
motion, and treat the carbons as an incoherent mixture of $2^4 = 16$ 
independent 3-spin systems, each with their resonance frequencies shifted 
by one of the $16$ possible sums of $\pm J_{\mathsf C, \mathsf H}/2$.
Because we start with the spins in the high-temperature equilibrium state, 
each of these $16$ independent evolutions contributes equally to the 
simulated density matrix.

The computer search for the strongly modulating control sequences
(pulses) is quite  demanding, and it is important to keep the
Hilbert space as small as possible during the associated simulations.
For this reason the simulations described in this paper ignored the
four hydrogen atoms in the alanine molecule, which were left aligned
with the main magnetic field during the actual experiments but which
nonetheless have couplings to the carbons on the order of $150$Hz.
A better way, which we subsequently implemented and is described
here for completeness, would be to average the results of $16$
simulations over each of the $3$-spin Hamiltonians
\begin{equation}
H_{\mathsf s, \mathsf c}^{\underline\delta} ~=~ \pi \sum_{i=1}^3 \Big( \nu_\mathsf C^i + \sum_{j=1}^4 \frac{{(-1)}^{\delta_j}}2\, J_{\mathsf C, \mathsf H}^{\,i,j} \Big) \sigma_{\mathsf C,\mathsf z}^i \:+\: \pi\! \sum_{j>i=1}^3 J_{\mathsf H, \mathsf H}^{\,i,j}\: \vec\sigma_{\mathsf H}^i \cdot \vec\sigma_{\mathsf H}^j ~,
\end{equation}
for all combinations of 
$\underline\delta \equiv [\delta_1 ,\ldots, \delta_4] = \{ 0, 1 \}^4$, 
where the ``$\mathsf c$'' stands for ``control''.
This is much faster than diagonalizing the Hamiltonian with the 
four hydrogens included ($16$ times longer as opposed to $\sim16^3$). 

During the free evolution periods between pulses, on the other hand, we 
applied additional (hard) $\pi$-pulses to the carbons to refocus this 
unwanted phase evolution, in accord with the free-evolution Hamiltonian used 
for these (much less demanding) simulations,
\begin{equation}
H_{\mathsf S, \mathsf f} ~=~ H_\mathsf s \:+\: \pi \sum_{i=1}^4 \sum_{j=1}^3 J_{\mathsf C, \mathsf H}^{\,i,j}\: \sigma_{\mathsf H,\mathsf z}^{\,i}\, \sigma_{\mathsf C,\mathsf z}^{\,j} ~,
\end{equation}
where the ``$\mathsf f$'' now stands for ``free''.
The frequency evolution of the hydrogens can be neglected because they are 
left along the $\mathsf z$-axis, thereby also avoiding population disturbances that 
would lead to nuclear Overhauser cross-relaxation and thereby memory effects.

The search for strongly modulating control sequences was further simplified 
by assuming that the RF phase, amplitude, and frequency was piecewise 
constant, so the total unitary transformation was given by the product
\begin{equation}
U_\mathsf{tot}^{\underline\delta} ~=~ \prod_{k=1}^{k_\mathsf{max}}\, U_k^{\underline\delta} ~=~ \prod_{k=1}^{k_\mathsf{max}}\, \mathrm T \exp\!\Big(\! -\mathrm i (t_k - t_{k-1}) H_{\mathsf S, \mathsf c}^{\underline\delta} \:-\: \mathrm i \int_{t_{k-1}}^{t_k} \mathrm dt\, H_{\mathsf{RF},\, k}(t) \Big)
\end{equation}
Here, ``$\mathrm T$'' denotes the usual time-ordering operator, and the 
external RF Hamiltonian during the $k$-th interval is given by
\begin{equation}
H_{\mathsf{RF},\,k}(t) ~=~ \alpha_k\, \gamma_\mathsf{C}\, B_1 \sum_{j=1}^3 \Big( e^{-\mathrm i (\pi t \nu_k + \phi_k ) \sigma_{\mathsf C,\mathsf z}^j}\, \sigma_{\mathsf C,\mathsf x}^{\,j}\, e^{\,\mathrm i (\pi t \nu_k + \phi_k ) \sigma_{\mathsf C,\mathsf z}^j} \Big) ,
\end{equation}
where $\alpha_k$ is the relative amplitude of the
field during the $k$th interval (cf.~Eqn.~\ref{eq:Hrf}).
This assumption allows the unitaries $U_k^{\underline\delta}$ for each 
interval to be calculated exactly by transforming to an interaction frame 
in which $H_{\mathsf{RF},\,k}$ becomes time-independent, and diagonalizing 
the net Hamiltonian in that frame \cite{E}. The spectrometer generates the 
control fields by applying a time-dependent voltage to a tuned resonator.
The limitations on both the control circuitry and the tuned resonator 
introduce time-dependent distortions of this modulation at the
discontinuities between intervals. To account for this limitation, we 
monitor the field generated in the control coil (the antenna that interacts 
with the spins) and pre-weight the time dependent wave-form to provide a 
close approximation to the desired shape.  In addition, we use the 
measured modulation sequence in the simulator to follow the dependence 
of the propagator for small distortions.

\begin{figure}[H] \newpage
\includegraphics[width=8.5cm]{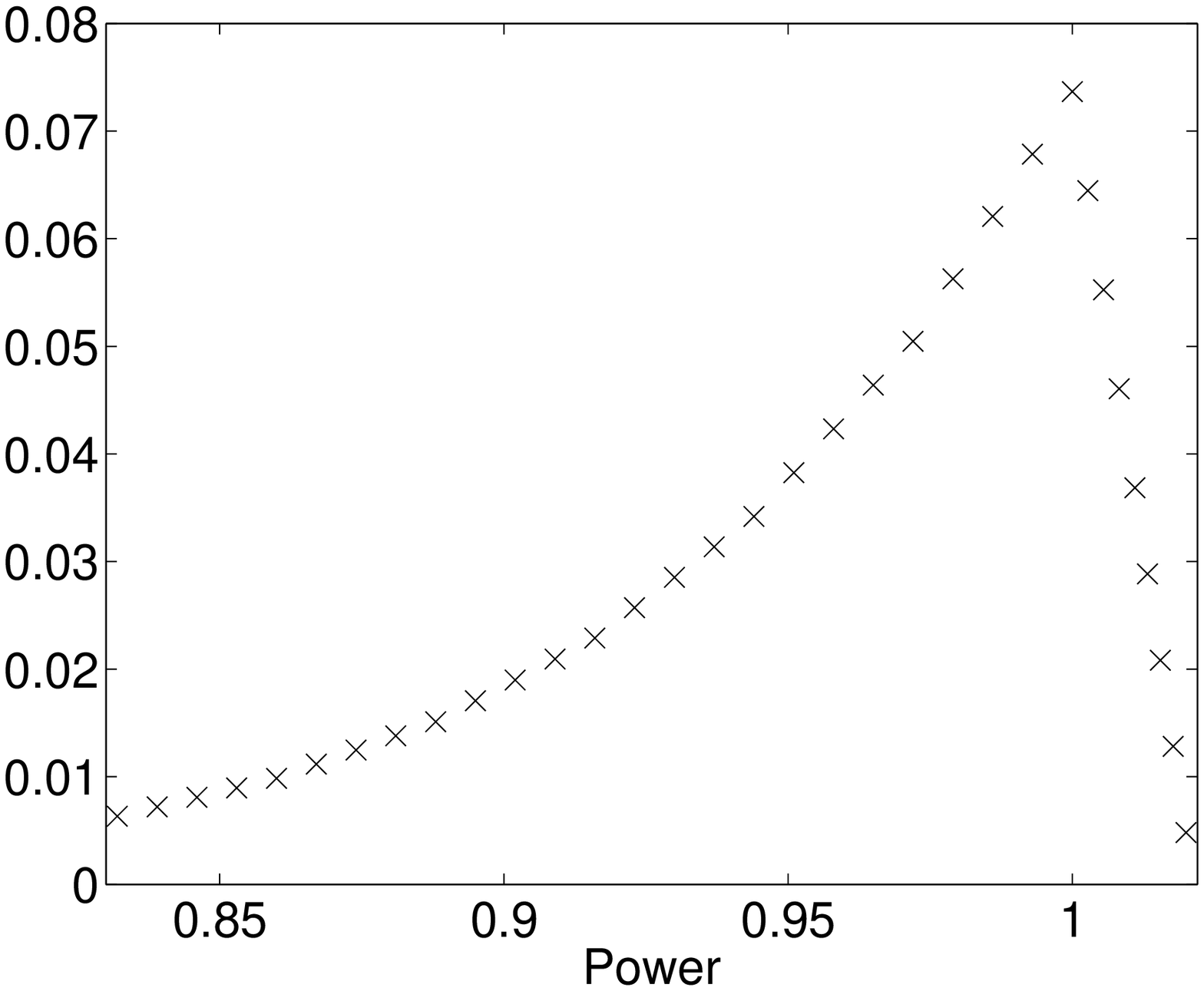}
\figcap{
The 33 point RF profile shows how much of the sample sees what fraction of
the desired power (so $1.0$ on the horizontal axis is the desired power).
This profile was measured experimentally and used for the simulations.
}\label{rfpdf}
\end{figure}

To achieve useable sensitivity, NMR is carried out on an ensemble of 
spatially distributed spins. The control field thus varies over the spatial 
extent of the sample, which is termed RF field inhomogeneity.
This variation could be reduced by using a larger coil, a smaller sample, 
or by selecting a particularly homogeneous region of the sample via a 
magnetic field gradient. All of these options, however, reduce the 
signal-to-noise ratio. Fortunately, the spatial variation of the RF 
field is constant over time, and can be measured very accurately.
This allows us to perform the simulations for each member of a histogram 
of the variations in RF amplitude across the sample, and to combine the 
results as an incoherent sum just as was described for the variations in the 
hydrogen spin states above. A total of $33$ values was included in this RF 
inhomogeneity histogram (see Fig.~\ref{rfpdf}).

The procedure used to find a modulation sequence that correctly
implements any desired unitary operation seeks to maximize
the fidelity between the desired unitary superoperator
$\bar U_\mathsf{th} \otimes U_\mathsf{th}$ and the
simulated superoperator $\mathcal S_\mathsf{op}$.
The latter was obtained as an incoherent or Kraus operator over the $33$ RF
field strengths in the experimentally measured RF inhomogeneity histogram, i.e.
\begin{equation}
\mathcal S_\mathsf{op}(\rho) ~=~ \sum_{\ell=1}^{33}\, p_\ell\, U_{\ell}\, \rho\, U_{\ell}^\dag ~\equiv~ \sum_{\ell=1}^{33}\, A_\ell\, \rho\, A_\ell^\dag ~.
\end{equation}
Thus, the gate fidelity as defined in Eqn.~(\ref{eqn:fid})
can be calculated directly from the Kraus operators $A_m$ as
\begin{eqnarray}
\hspace{-2em} F_\mathsf{e}\big( (\bar U_\mathsf{th} \otimes U_\mathsf{th})^\dag \mathcal S_\mathsf{op} \big) &\equiv& \frac{\mathrm{tr}\big( (\bar U_\mathsf{th} \otimes U_\mathsf{th})^\dag \mathcal S_\mathsf{op} \big)}{\mathrm{tr}\big((\bar U_\mathsf{th}^\dag\bar U_\mathsf{th})\otimes(U_\mathsf{th}^\dag U_\mathsf{th}) \big)} ~=~ \frac1{64} \sum_{m=1}^{33} \mathrm{tr}\big( (\bar U_\mathsf{th} \otimes U_\mathsf{th})^\dag (\bar A_m \otimes A_m) \big) \\ \nonumber
&=& \frac1{64} \sum_{m=1}^{33} \mathrm{tr}\big( (\bar U_\mathsf{th}^\dag \bar A_m) \otimes (U_\mathsf{th}^\dag A_m ) \big) ~=~ \frac1{64} \sum_{m=1}^{33} \big| \mathrm{tr}( U_\mathsf{th}^\dag A_m ) \big|^2 ~.
\end{eqnarray}
In simulations including the four hydrogen spins,
the sum on the right-hand side of this formula must be increased by
another factor of $16$ in taking the partial trace over the hydrogen spins.

\begin{figure}[H] \newpage
\includegraphics[width=8.5cm]{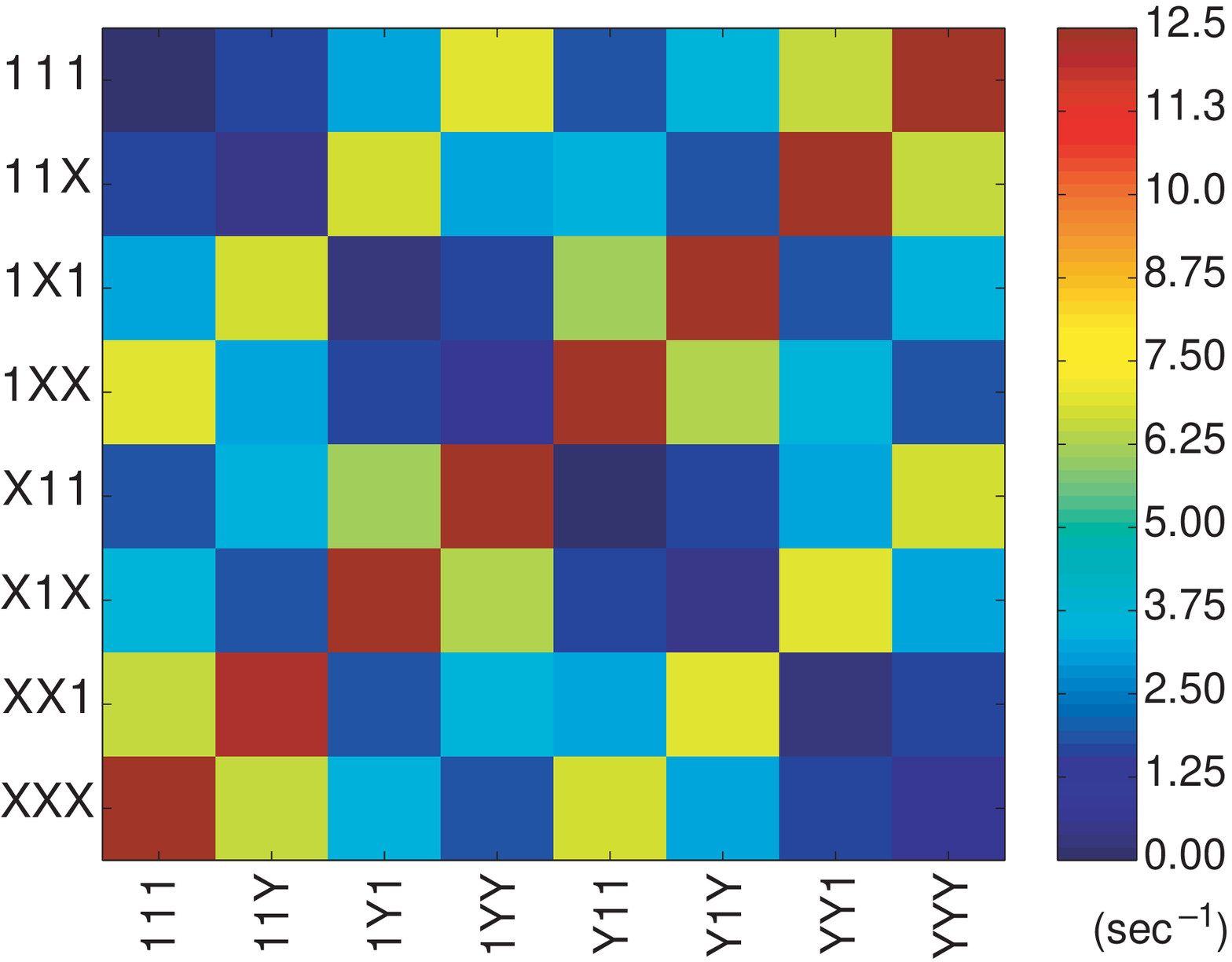}
\figcap{Measured Hadamard relaxation operator versus the product operator basis for the three carbons of alanine in the absence of RF fields, with rates color-coded as indicated by the legend. The labels on the axes are a short-hand for the product operators associated with with $\mathsf x$ and $\mathsf y$ directions on the three carbons, with e.g.~$\mathsf{X1X} \leftrightarrow \sigma_\mathsf x^1 \sigma_\mathsf x^3$. The labels for each entry of the matrix are obtained by analogy with the product rules for the Pauli operators; for example, the matrix entry in the row labeled~$\mathsf{1XX}$ and column labeled $\mathsf{Y1Y}$ is $\mathsf{YXZ} \leftrightarrow \sigma_\mathsf y^1 \sigma_\mathsf x^2 \sigma_\mathsf z^3$ \cite{Tim2}. See Table \ref{tab:rel} for the numerical values of the various entries.
} \label{fig:had}
\end{figure}

The relaxation superoperator for the carbons in alanine
was measured in the absence of RF fields, and found to have
$98.5$\% of its norm along the diagonal in the product operator basis.
This means that the various product operator components decay 
mono-exponentially, without  cross-relaxation, so that it can be 
described by an $8\times8$ ``Hadamard relaxation matrix'' \cite{Tim2,Tim3}.
This is shown pictorially in Fig.~\ref{fig:had}, while precise values for
the various types of rates seen in the figure are given in Table \ref{tab:rel}.

\begin{table}[tb]
\caption{ \label{tab:rel} } \medskip
\parbox{0.75\textwidth}{\setlength{\baselineskip}{0.6\baselineskip}
Measured carbon-13 decay rates for each type of product operator (s$^{-1}$)
in the alanine molecule used for the experiments. These include the inverses
of the relaxation times $T_1$ and $T_2$ for all three carbon-13 spins. 
The ``$\mathsf x|\mathsf y$'' subscripts indicate that,
in every operator of the corresponding product, all the labels
to the left or to the right of the bar must be used together,
thus indicating a total of two products with one expression.
\bigskip}
\begin{tabular}{||l|c||l|c||} \hline\hline
\textsf{Type of Product Operator} & \textsf{Decay Rate} &
\textsf{Type of Product Operator} & \textsf{Decay Rate}
\\ \hline\hline
$~\sigma_\mathsf z^1$ & $0.032\,\pm\,0.01$ & $~\sigma_\mathsf z^2$ & $0.345\,\pm\,0.01$ \\\hline
$~\sigma_\mathsf z^3$ & $0.583\,\pm\,0.01$ & $~\sigma_\mathsf{z}^1 \sigma_\mathsf{z}^2$ & $0.282\,\pm\,0.01$ \\\hline
$~\sigma_\mathsf{z}^1 \sigma_\mathsf{z}^3$ & $0.458\,\pm\,0.01$ & $~\sigma_\mathsf z^2  \sigma_\mathsf z^3$ & $0.689\,\pm\,0.01$ \\\hline
$~\sigma_\mathsf z^1\sigma_\mathsf z^2 \sigma_\mathsf z^3$ & $0.684\,\pm\,0.01$ & $\begin{array}{l} \sigma_\mathsf{x\smash|y}^1,\; \sigma_\mathsf{x\smash|y}^1\sigma_\mathsf z^2, \\[-0.5ex] \sigma_\mathsf{x\smash|y}^1\sigma_\mathsf z^3,\;  \sigma_\mathsf{x\smash|y}^1\sigma_\mathsf z^2\sigma_\mathsf z^3 \end{array}$ & $1.89\,\pm\,0.02$ \\\hline
$\begin{array}{l} \sigma_\mathsf{x\smash|y}^2,\; \sigma_\mathsf z^1 \sigma_\mathsf{x\smash|y}^2, \\[-0.5ex] \sigma_\mathsf{x\smash|y}^2\sigma_\mathsf z^3,\;  \sigma_\mathsf z^1\sigma_\mathsf{x\smash|y}^2\sigma_\mathsf z^3 \end{array}$ & $3.19\,\pm\,0.01$ & $\begin{array}{l} \sigma_\mathsf{x\smash|y}^3,\; \sigma_\mathsf z^1\sigma_\mathsf{x\smash|y}^3, \\[-0.5ex] \sigma_\mathsf z^2\sigma_\mathsf{x\smash|y}^3,\;  \sigma_\mathsf z^1\sigma_\mathsf z^2 \sigma_\mathsf{x\smash|y}^3 \end{array}$ & $1.68\,\pm\,0.01$ \\\hline
$\begin{array}{l} \sigma_\mathsf{x\smash|y}^1 \sigma_\mathsf{y|x}^2,\; \sigma_\mathsf{x\smash|y}^1 \sigma_\mathsf{y\smash|x}^2 \sigma_\mathsf z^3, \\[-0.5ex] \sigma_\mathsf{x\smash|y}^1 \sigma_\mathsf{x\smash|y}^2,\; \sigma_\mathsf{x\smash|y}^1 \sigma_\mathsf{x\smash|y}^2  \sigma_\mathsf z^3 \end{array}$ & $6.93\,\pm\,0.11$ & $\begin{array}{l} \sigma_\mathsf{x\smash|y}^1 \sigma_\mathsf{y\smash|x}^3,\; \sigma_\mathsf{x\smash|y}^1 \sigma_\mathsf z^2 \sigma_\mathsf{y\smash|x}^3, \\[-0.5ex] \sigma_\mathsf{x\smash|y}^1 \sigma_\mathsf{x\smash|y}^3,\; \sigma_\mathsf{x\smash|y}^1 \sigma_\mathsf z^2 \sigma_\mathsf{x\smash|y}^3 \end{array}$ & $3.56\,\pm\,0.06$ \\\hline
$\begin{array}{l} \sigma_\mathsf{x\smash|y}^2 \sigma_\mathsf{y|x}^3,\; \sigma_\mathsf z^1 \sigma_\mathsf{x\smash|y}^2\sigma_\mathsf{y|x}^3, \\[-0.5ex] \sigma_\mathsf{x\smash|y}^2 \sigma_\mathsf{x\smash|y}^3,\; \sigma_\mathsf z^1 \sigma_\mathsf{x\smash|y}^2\sigma_\mathsf{x\smash|y}^3 \end{array}$ & $6.81\,\pm\,0.10$ & $\begin{array}{l} \sigma_\mathsf{y\smash|x}^1 \sigma_\mathsf{x\smash|y}^2 \sigma_\mathsf{x\smash|y}^3,\;  \sigma_\mathsf{x\smash|y}^1 \sigma_\mathsf{y\smash|x}^2 \sigma_\mathsf{x\smash|y}^3, \\[-0.5ex]  \sigma_\mathsf{x\smash|y}^1 \sigma_\mathsf{x\smash|y}^2 \sigma_\mathsf{y\smash|x}^3 \end{array}$ & $13.48\,\pm\,0.81$ \\\hline
$~\sigma_\mathsf{x\smash|y}^1 \sigma_\mathsf{x\smash|y}^2 \sigma_\mathsf{x\smash|y}^3$ \hspace{2.2cm} & $14.58\,\pm\,0.20$ &&\\\hline\hline
\end{tabular}
\end{table}

To be complete, the relaxation superoperator should also be measured as a function
of the applied RF fields (and include memory effects due to nuclear Overhauser 
effects with the hydrogens \cite{Ernst}). The QFT, however, efficiently mixes the
states of the  three carbon spins, so that on average over its implementation the 
decoherence is indistinguishable from a uniform, isotropic attenuation of all the 
product operator components. That is to say, QPT on the QFT is not able to provide 
any details regarding the physical relaxation processes operative in alanine, 
save for an average overall rate of attenuation. Of course, QPT could be 
performed on a simpler gate, as in Ref.~\cite{Nick} where it was used to 
derive a relaxation superoperator for free precession, but the goal here is 
to use QPT to learn about the coherent and incoherent errors committed during 
implementation of a complex unitary transformation, the QFT. Therefore, this 
simple effective relaxation superoperator was assumed to accelerate the 
simulations of the overall QFT (no relaxation was assumed during the pulse 
design simulations, since they are not intended to correct for such effects).

\section{An Overview of the Experimental Results}
\begin{figure}[H] \newpage
\includegraphics[width=8.5cm]{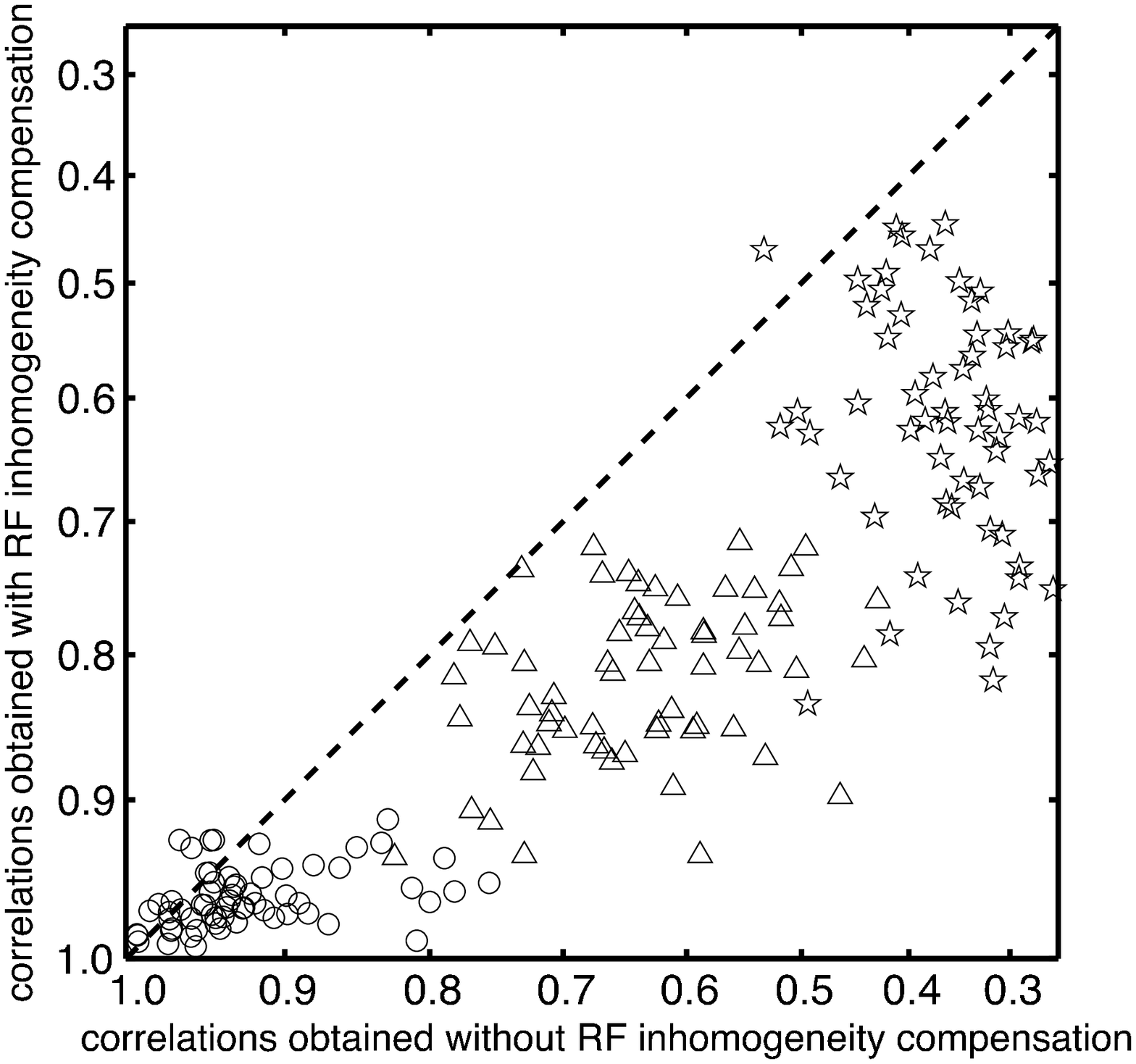}
\figcap{
The various correlations (see Section \ref{sec:qpt}) for the $64$ states of the product
operator basis before and after applying the QFT, where those obtained from the
uncompensated strongly modulating control sequences have been plotted against those
obtained with the sequences designed to compensate for RF inhomogeneity \cite{Marco}.
To spread the points out, the (co)ordinates plotted are $\log_{10}(1 + (1 - x))$,
although the axes have been labeled by the actual correlations $0\le x \le 1$.
Circles (\protect\raisebox{1pt}{$\scriptscriptstyle\bigcirc$})
are used for the initial state correlations, while triangles
($\scriptstyle\triangle$) denote the final state correlations
and stars (\protect\raisebox{1pt}{$\scriptstyle\bigstar$})
indicate the final state attenuated correlations.
For the uncompensated sequences the average initial state
correlation is 0.93, the average final state correlation
is 0.64, and the  average attenuated correlation is 0.37. 
For the compensated sequences, the average initial state correlation is 0.96, 
the average final state correlation is 0.82, and the gate fidelity 0.64.
Moreover, since a large majority of the points lie below the
diagonal in the plot (dashed line), the correlations with the
compensated pulses were in fact better in almost every case.
} \label{metrics}
\end{figure}

\begin{figure}[H] \newpage
\includegraphics[width=7cm]{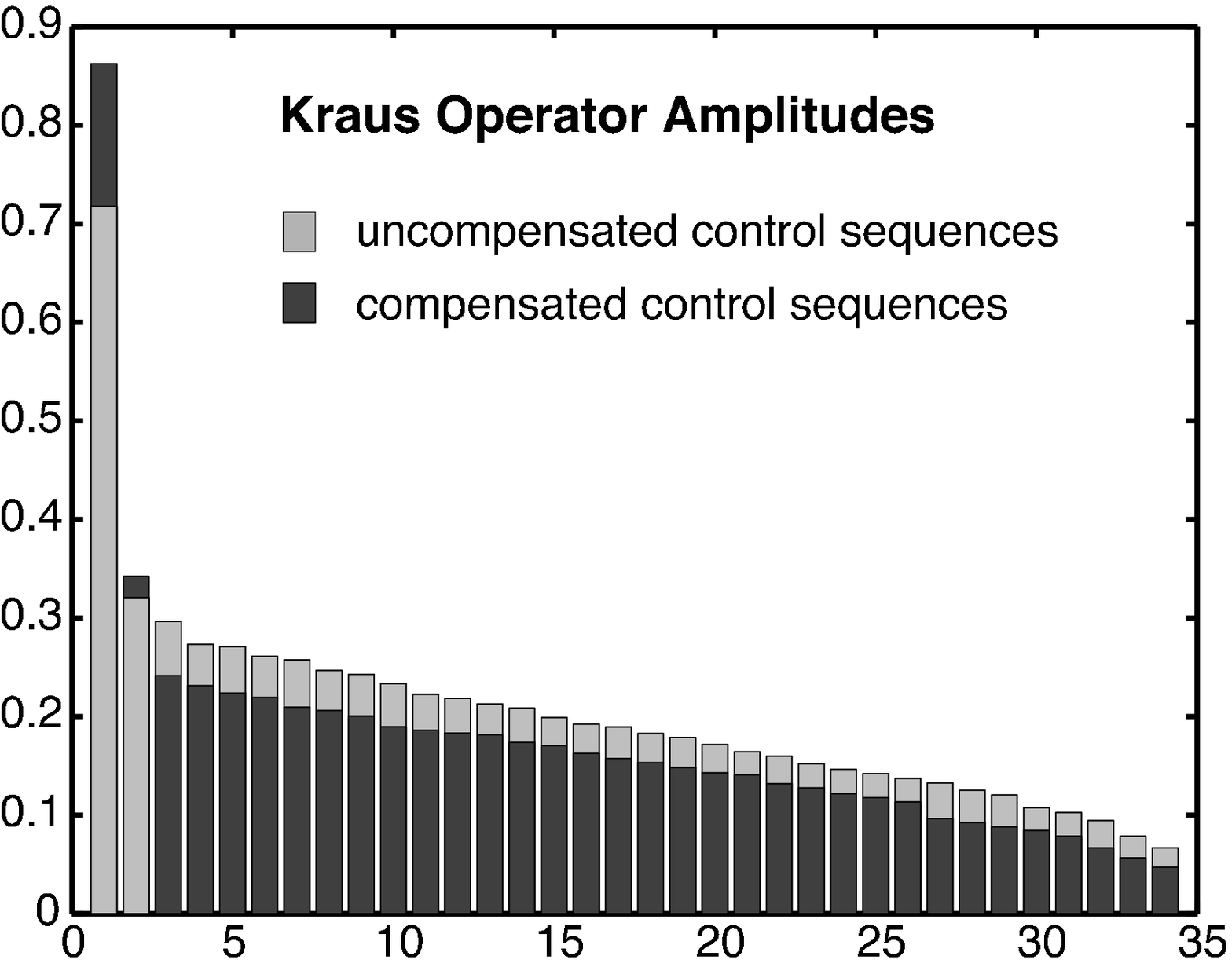} \hspace{0.5cm}
\includegraphics[width=8.5cm,height=6.5cm]{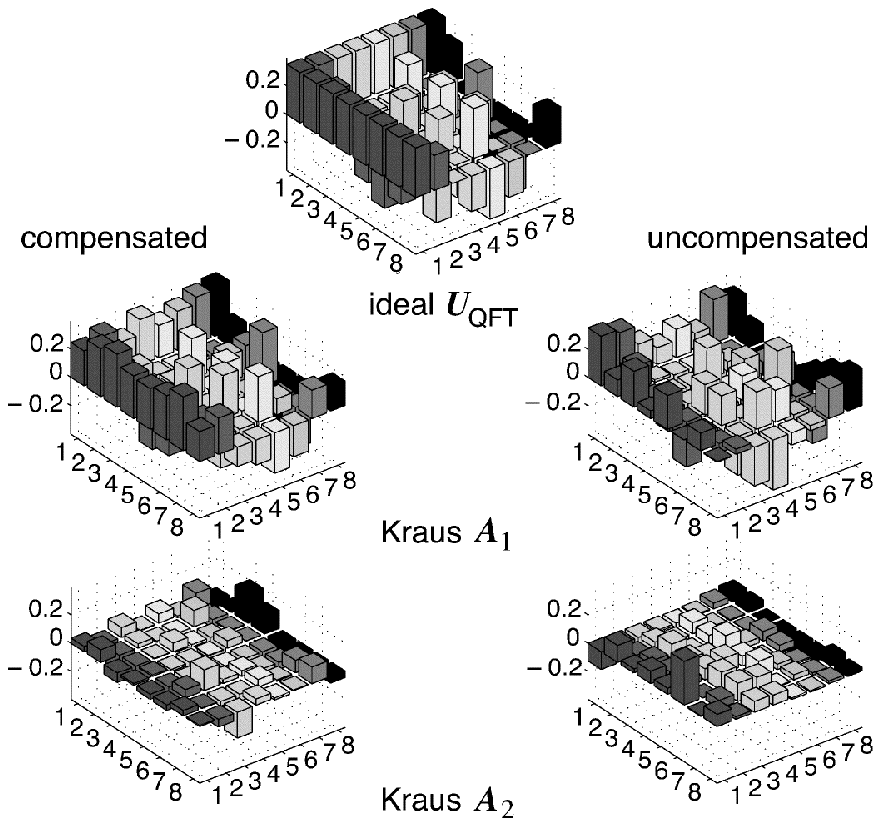}
\figcap{
The bar graph on the left shows the amplitudes $a_k = \|A_k\|/\sqrt8$
of experimental QFT Kraus operators, where the light grey
portions of the bars are the amplitudes obtained from
QPT done with the uncompensated control sequences,
and the dark grey bars those obtained from QPT done
with the RF-compensated control sequences (see text).
With the uncompensated control sequences the dominant
Kraus operator had an amplitude $a_1^\mathsf u = 0.72$,
whereas the compensated had $a_1^\mathsf c = 0.86$.
The second largest Kraus operators had amplitudes of
$a_2^\mathsf u = 0.30$ and $a_2^\mathsf c = 0.34$,
respectively, after which the amplitudes decreased much more slowly.
The plots on the right (which are shaded by column number to aid
viewing) compare the real part of the desired unitary operator
$U_\mathsf{QFT}$ (top center) to the real part of the largest Kraus
operator $A_1$ obtained using the uncompensated and compensated control
sequences (left and right-hand plots on the second row, respectively),
and to the corresponding second largest Kraus operators $A_2$ (bottom row).
The correlations between real parts of these operators were
$C(\Re U_\mathsf{QFT},\Re A_1^\mathsf u) = 0.78$,
$C(\Re U_\mathsf{QFT},\Re A_1^\mathsf c) = 0.95$,
$C(\Re U_\mathsf{QFT},\Re A_2^\mathsf u) = 0.06$ and
$C(\Re U_\mathsf{QFT},\Re A_2^\mathsf c) = 0.02$.
} \label{OldNewKrAmp}
\end{figure}

Complete QPT was performed twice using different sets of strongly modulating
RF control sequences to implement single spin rotations (Section \ref{sec:exp}).
The first such iteration was done with the sequences described
in Ref.~\cite{E}, which refocussed the evolution under the
alanine molecule's internal spin Hamiltonian ($H_\mathsf s$
in Eqn.~\ref{H_small_s}) during the sequence.
The second iteration used control sequences that not only
refocussed the internal Hamiltonian, but also compensated
for inhomogeneity in the RF field itself \cite{Marco}.
Thus, the second set of control sequences was expected
to produce much smaller incoherent errors than the first,
while the coherent and decoherent errors
were expected to be roughly the same for both.
The improvement obtained with the RF-compensated set is
illustrated in Fig.~\ref{metrics}, which plots the correlations
and attenuated correlations obtained for each of the 64 product operator
basis states used as inputs for the two iterations against one another.
Further evidence for substantial improvements with the RF-compensated
control sequences may be obtained from the Kraus operator plots and
statistics in Fig.~\ref{OldNewKrAmp} (see below for their interpretation).
Having demonstrated this clear-cut improvement, all subsequent analysis will
be given only for the results obtained wiht the compensated control pulses.

The correlation provides an estimate of the accuracy of
the experimental QFT implementation, but without considering
the loss of magnetization due to decoherence or incoherence.
The average of the input state correlations over all 64 basis states was $0.96$,
while the average correlation following application of the QFT was $0.82$.
A minimization search, however, found pure states which, after operating
on them with the experimental superoperator, had a correlation with
the same pure state following the theoretical QFT as low as $0.45$.
The average attenuated correlation (or gate fidelity \cite{E}),
which also takes into account the loss of magnetization, was $0.64$,
indicating that about $22$\% of the magnetization was lost over the
ca.~30 ms needed to implement the QFT (see Section \ref{sec:exp}).
These numbers are in-line with expectations based on other recent
applications of the RF-compensated control sequences \cite{Swap}.
About half of this magnetization loss was expected due to intrinsic
decoherence, and since approximately the same amount of decoherence
occurred during the input states' readout as during the output states',
the remainder is probably due to residual uncompensated incoherence and/or
imperfect decoupling of the protons from the carbons used as qubits.
These issues will be discussed in more detail in later sections.
A complete list of all the output state (attenuated)
correlations may be found in Fig.~\ref{AllCorBarH} below.

Rather than looking at the action of the QFT on states, one can also look
directly at the Kraus operator sum computed from the completely positive part
of the experimental supermatrix $\mathcal M_\mathsf{obs}$ (Section \ref{sec:incoh}).
Each Kraus operator $A_k$ has an associated amplitude,
$a_k = \|A_k\|/\sqrt8 = \sqrt{\lambda_k/8}$
(where $\lambda_k$ is the $k$th eigenvalue of the associated Choi supermatrix),
and since in a perfect implementation the desired unitary operator
$U_\mathsf{QFT}$ would be the only Kraus operator with nonzero amplitude,
one expects the Kraus operator with the largest amplitude
to be at least fairly similar to $U_\mathsf{QFT}$.
This is confirmed by the plots of the real part of the
largest Kraus operator, shown in Fig.~\ref{OldNewKrAmp},
which had an amplitude $a_1 = 0.86$ anda correlation
with the real part $U_\mathsf{QFT}$ of 0.95,
implying a net coherent error of roughly 5\%.
The second largest Kraus operator, also shown, had an amplitude of $a_2 = 0.34$,
and was also rather close to unitary although uncorrelated with $U_\mathsf{QFT}$.
We expect it to be a rough approximation to the largest unitary operator
in any sum of unitary transformations making up the incoherent error. 
Finally, there are another 32 essentially non-unitary Kraus operators
in the completely positive part of the experimental QFT supermatrix
with smaller amplitudes (Fig.~\ref{OldNewKrAmp}).

The non-completely positive part of the QFT supermatrix,
which is obtained from the eigenvectors associated with
the negative eigenvalues of its Choi supermatrix,
will also be of interest in what follows.
The positivity of the QFT supermatrix, as defined
at the end of Section \ref{sec:incoh}, was only $0.60$,
but the ratio of the smallest to the largest Choi supermatrix
eigenvalues was $-0.075$, indicating that the negative eigenvalues
were rather small in magnitude in comparison to the positive.
The ratio with the second largest was $-0.48$, and the
third had almost the same magnitude as the smallest.
These observations, together with the well-known sensitivity
of the eigenvectors of nearly degenerate eigenvalues to
small perturbations in the elements of the matrix from
which they come, imply that rather little information about
the errors made in the QFT implementation can be gleaned
from the individual Kraus operators after the first two.
Only the subspaces spanned by the eigenvectors associated
with all of the smaller positive, or perhaps negative,
eigenvalues is likely to be statistically significant.

\section{Comparison with the Model} \label{sec:comp}
To assess the precision and completeness of our
system-plus-apparatus model, the complete set of experiments
involved in QPT on the QFT was simulated using the mathematical
model of the system-plus-apparatus described in Section \ref{sec:model}.
The compatibility of the simulated results with the
experimental provides a rigorous test for the accuracy of the model,
which takes all the known significant imperfections
of the experimental apparatus into account.
All of the unitary operations performed during these simulations were 
implemented using exactly the same strongly modulating control sequences
that were used for the experiments, and complete state tomography was 
performed for each input and output state. It should be noted, however,
that these states were reconstructed directly from the observables
in the simulated density matrices, without further simulating the
spectra and fitting them as required in experimental state tomography
(which of course gives rise to additional errors in the actual experiments). 
Figure \ref{AllCorBarH} shows the correlation and attenuated correlation 
of the initial and final states obtained from these simulations.

Figure \ref{ExpVsSimKrAmp} plots the sorted Kraus operator amplitudes
obtained from the simulated supermatrix, along with those from the
experimental supermatrix for comparison (cf.~Fig.~\ref{OldNewKrAmp}).
The negative values plotted are actually the negative square roots
of the corresponding Choi matrix eigenvalues, and are shown to
illustrate that the experimental supermatrix was significantly
further from being completely positive than was the simulated
(the positivity of the simulated was $0.86$,
as opposed to $0.60$ for the experimental).
The most likely reason for this is the absence in the simulations
of the additional errors expected from fitting the spectra to
extract the product operator amplitudes for state tomography.
Also shown once again is the real part of the theoretical
QFT unitary matrix (cf.~Fig.~\ref{OldNewKrAmp}),
along with the real parts of the matrices of the largest
Kraus operators from the experimental and simulated supermatrices.
Finally, the corresponding best unitary approximations to the largest
Kraus operators, obtained by setting their singular values to unity,
have their real parts shown on the bottom right-hand side of Fig.~\ref{ExpVsSimKrAmp}.
It may be seen that there is a good correspondence between the largest
Kraus operators as well as between their best unitary approximations, with a
correlation between the simulation and the experiment of $0.90$ in both cases.
As noted previously, the smaller Kraus operators cannot
be expected to correspond significantly to one another.

Since the Fourier transformation converts between the position and
momentum bases, it has a very simple interpretation in phase space.
Classically, a Fourier  transform rotates phase space by $90^\circ$.
In quantum phase space, the QFT superoperator has only one
nonzero element equal to unity in each row and column,
and so constitutes a permutation matrix \cite{CPS}.
This is shown in Fig.~\ref{PhSp}, along with the corresponding
plots for the simulated and experimental supermatrices.
One can see immediately from these plots that
there are errors in the QFT implementation.
The phase space basis, however, consists of
operators that are neither Hermitian nor
products of any underlying Hilbert space basis,
and which are not related in any simple way to
the physical operators used to implement the QFT.
Thus, although the action of the QFT is very
easy to understand in the phase space basis,
it is very hard to interpret the discrepancies
between the theoretical superoperator and observed supermatrix
in terms of implementation errors in this basis.
It is worth noting, nonetheless, that in the phase space basis
the errors appear more white than Gaussian especially in the simulation.

\begin{figure}[H] \newpage
\includegraphics[height=4.75cm]{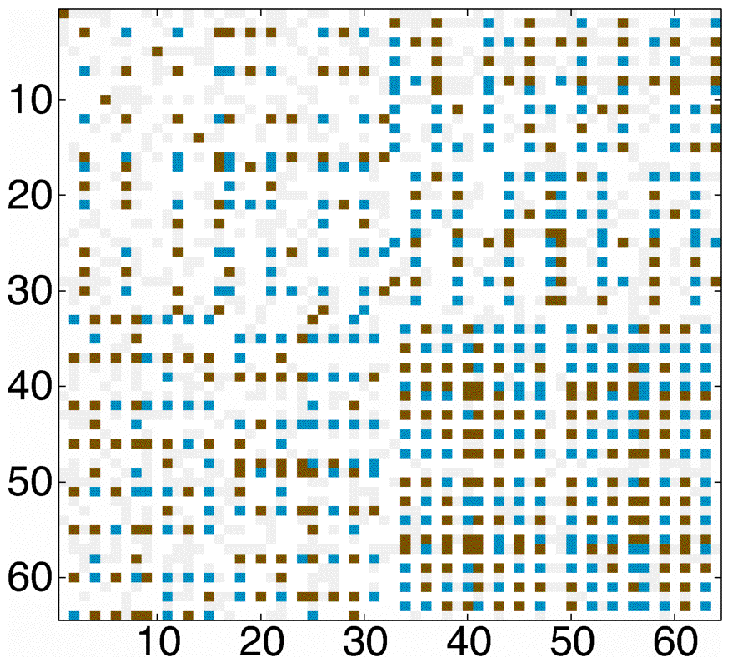}
\includegraphics[height=4.75cm]{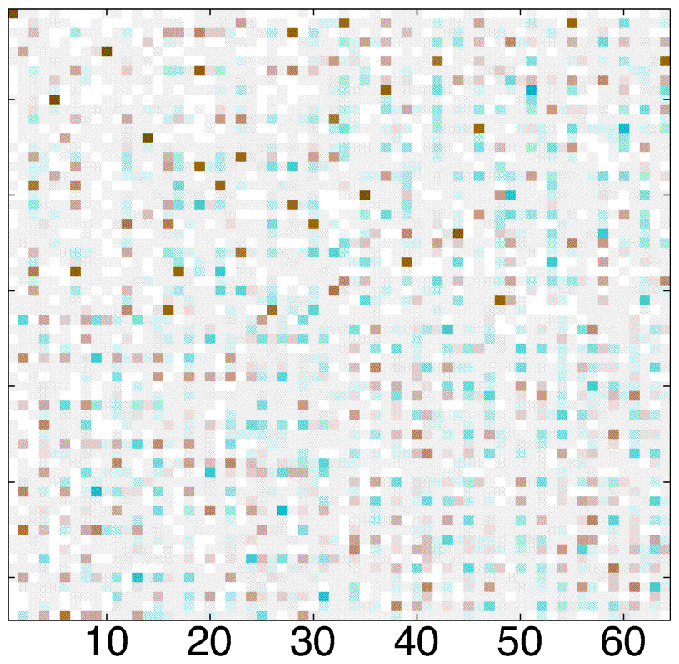}
\includegraphics[height=4.85cm]{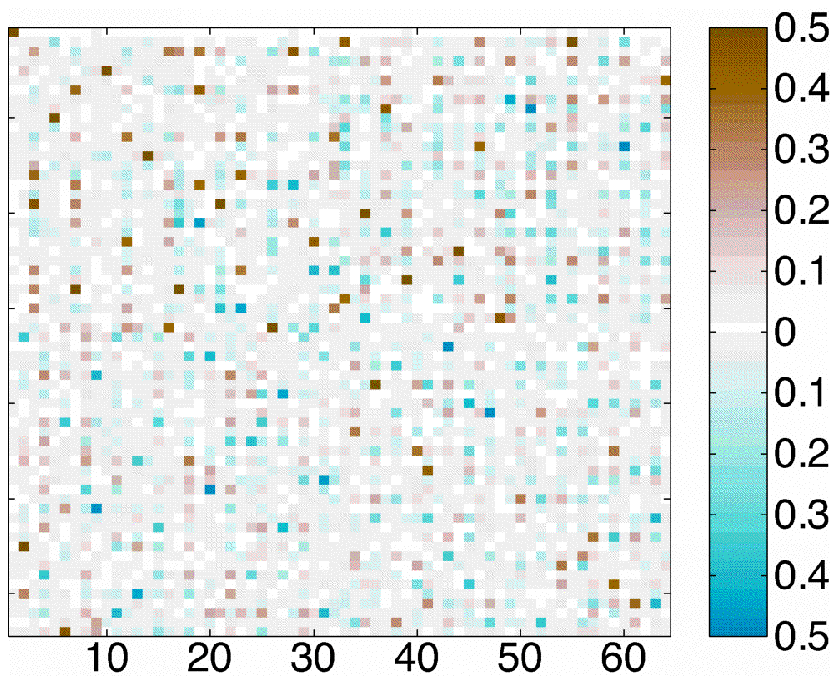}
\figcap{
The theoretical (left), simulated (middle) and experimental supermatrices
in the product operator basis, wherein all the elements are necessarily real
(see text; the scale has been reduced by a factor of $2$ to create more contrast).
The correspondence between the rows / columns of this supermatrix and
the individual product operators is the same as that given by the
labels on the vertical axis of the bar graph in Fig.~\ref{AllCorBarH}.
Because all the entries of the corresponding ``real'' density matrix are
the expectation values of observables which transform nicely under rotations
\cite{Tim2}, the physical interpretation is somewhat easier in this basis.
This is manifest, for example, in the appearance of several fixed points
with a $1$ on the diagonal and zeros elsewhere in the same row (see text).
} \label{POba}
\end{figure}

\begin{figure}[H] \newpage
%\parbox{16cm}{
\includegraphics[angle=270,width=15cm]{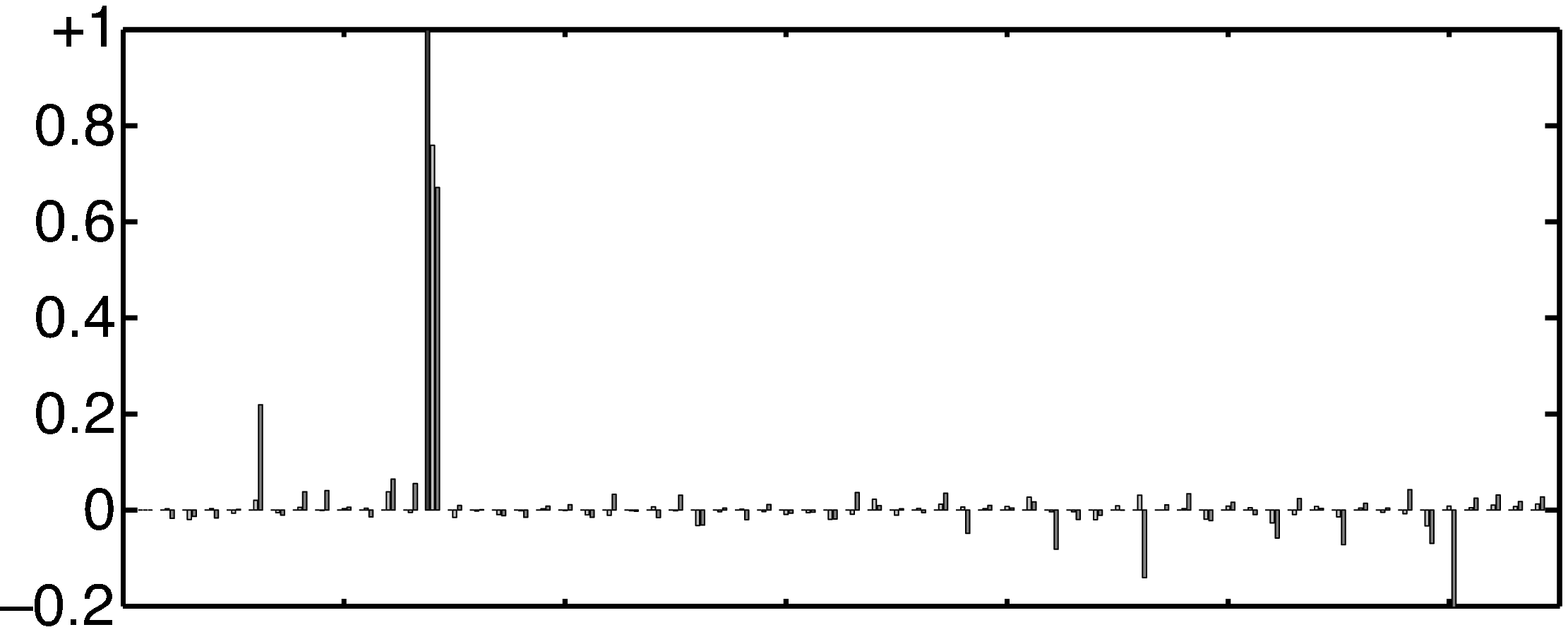}
\includegraphics[angle=270,width=15cm]{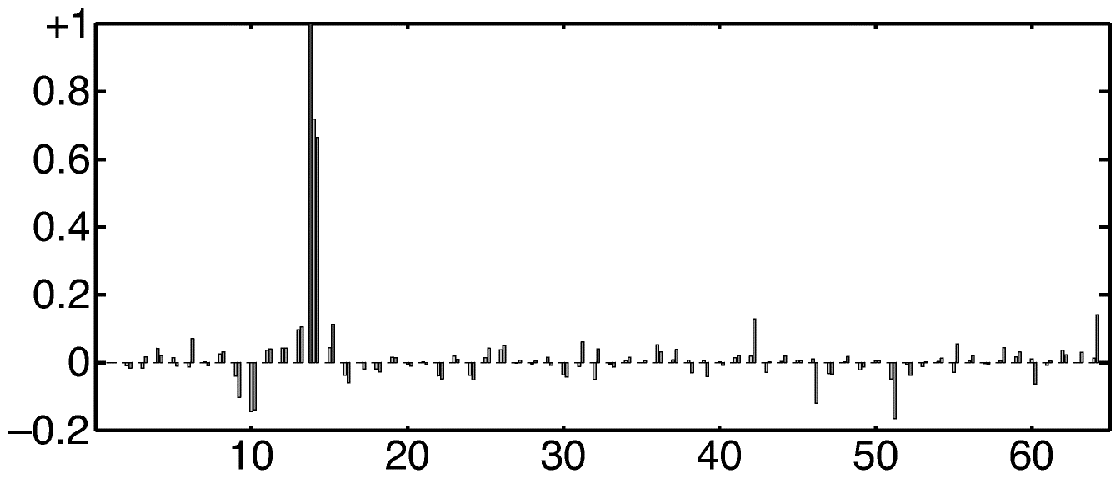}
%} %\renewcommand{\baselinestretch}{1.0}
\figcap{
As may be seen from row 14 of the supermatrices shown in Fig.~\ref{POba},
the operator $\sigma_\mathsf x^1 \sigma_\mathsf z^3$ is a fixed point of the QFT.
The upper bar graph displays the corresponding row of the theoretical (dark grey),
simulated (medium grey) and simulated with coherent errors corrected (light grey)
supermatrix elements versus the product operator basis, whereas the lower
bar graph displays these same statistics for the experimental supermatrix.
In contrast to the phase space and Zeeman basis (not shown),
there is a significant correlation between the experimental
and simulated values; specifically, the simulated-to-experimental
correlation coefficients were $0.80$ and $0.94$
before and after correction, respectively.
In the present case these correlations are due almost
entirely to the single large diagonal value, but similar
correlations were obtained for all other pairs of rows;
the average correlations were $0.78$ and $0.94$
before and after correction, respectively.
} \label{POrows}
\end{figure}

Other bases, in which the discrepancies between the simulated
and experimental supermatrices are also clearly manifested,
include the computational basis (or Zeeman basis; see Fig.~\ref{Zeem}), 
and the product operator basis (shown in Fig.~\ref{POba}), where the latter 
consists of all possible products of the Pauli matrices $\sigma_\mathsf{x}$, 
$\sigma_\mathsf{y}$ and $\sigma_\mathsf{z}$ of different spins \cite{Ernst}
(see Section \ref{sec:qpt} above). It may be observed that the theoretical 
superoperator contains several fixed points in the product operator basis,
which provide us with another interesting metric for the precision of our 
implementation. Two of these fixed points are
$\sigma_\mathsf{x}^1\sigma_\mathsf{z}^3$ and $(\sigma_\mathsf{x}^1+\sigma_\mathsf{z}^3)/2$.
The correlations between these fixed points, before and after applying the
experimental and simulated superoperators, are 0.91 and 0.93 respectively.
Further information can be obtained from plots of the individual rows,
which depict how much each input state contributes to a given
output state, as shown in the relatively simple case of the
$\sigma_\mathsf{x}^1\sigma_\mathsf{z}^3$ fixed point in Fig.~\ref{POrows}.

The greatest part of the deviation between the simulated and experimental
results is expected to be due to the propagation of small coherent errors.
Thus it is interesting to compare the simulated and experimental
supermatrices after correcting for these errors as completely as possible.
This is most simply done by left-multiplying by the inverse of the superoperator
obtained from the best unitary approximation to the largest Kraus operator,
$\overline U_1^\dag \otimes U_1^\dag$, followed by the theoretical unitary
superoperator $\overline U_\mathsf{QFT} \otimes U_\mathsf{QFT}$, i.e.
\begin{equation}
\mathcal M_{\mathsf{cor},\mathsf{obs|sim}} ~=~
\big( \overline{U}_\mathsf{QFT} \otimes U_\mathsf{QFT} \big)
\big( \overline U_{1,\mathsf{obs|sim}} \otimes U_{1,\mathsf{obs|sim}} \big)^\dag\,
\mathcal M_{\mathsf{obs}|\mathsf{sim}} ~.
\end{equation}

The resulting supermatrices in the phase space basis
are displayed in Fig.~\ref{CorrectedSuperPSB}, along with
the exact QFT supermatrix in the same basis for comparison.
It may immediately be seen that there has been a considerable improvement
in the similarity of these matrices, as is further confirmed by a
correlation between the corrected simulated and corrected experimental
supermatrices of $0.94$, between the corrected simulated and theoretical of $0.99$,
and between the corrected experimental and theoretical of $0.95$.
The correlations between the simulated and experimental correction factors,
i.e.~the products of the unitary superoperators in the above equation,
are however only $0.90$, indicating that the cumulative effects
of coherent errors that were not taken into account by the
simulations over the course of the experiments was roughly $10$\%.
It should further be noted that whereas the experimental supermatrix
is significantly more strongly correlated with the theoretical than
it is with the simulated, after they have been corrected their
correlations and attenuated correlations with the theoretical
are very nearly the same.

\begin{table}[tb]
\renewcommand{\baselinestretch}{1.0}
\caption{ \label{tab:allcor} \bigskip}
\parbox{0.91\textwidth}{\setlength{\baselineskip}{0.6\baselineskip}
The correlation coefficients between all pairs of supermatrices, labelled as described in the main text, together with the attenuated correlations (as defined in section II) between each and the theoretically exact supermatrix of the quantum Fourier transform.
\bigskip}
\begin{tabular}{||@{\hspace{12pt plus 6pt minus 6pt}}
l@{\hspace{12pt plus 6pt minus 6pt}}||
*{5}{@{\hspace{12pt plus 6pt minus 6pt}}
c@{\hspace{12pt plus 6pt minus 6pt}}}|c||}
\hline\hline
Supermatrix & \multicolumn{5}{c|}{\hspace{-12pt}Correlations Among All Pairs of Supermatrices\ } & \rule{0pt}{3.5ex}
\raisebox{3pt}{\parbox{0.9in}{\baselineskip=12pt Attenuated Correlation}} \\
\hline\hline
Theoretical & 1.00 & 0.89 & 0.99 & 0.82 & 0.95 & 1.00 \\
Simulated & -- & 1.00 & 0.90 & 0.79 & 0.85 & 0.68 \\
Corrected Sim. & -- & -- & 1.00 & 0.81 & 0.94 & 0.76 \\
Experimental & -- & -- & -- & 1.00 & 0.87 & 0.64 \\
Corrected Exp. & -- & -- & -- & -- & 1.00 & 0.74 \\
%Simulated & 0.89 & 0.68 \\
%Experimental & 0.82 & 0.64 \\
\hline\hline
\end{tabular} \\\bigskip
\leftline{\hspace{0.45in}}
\end{table}

\section{Discussion} \label{sec:disc}
\begin{figure}[H] \newpage
\includegraphics[width=8.5cm]{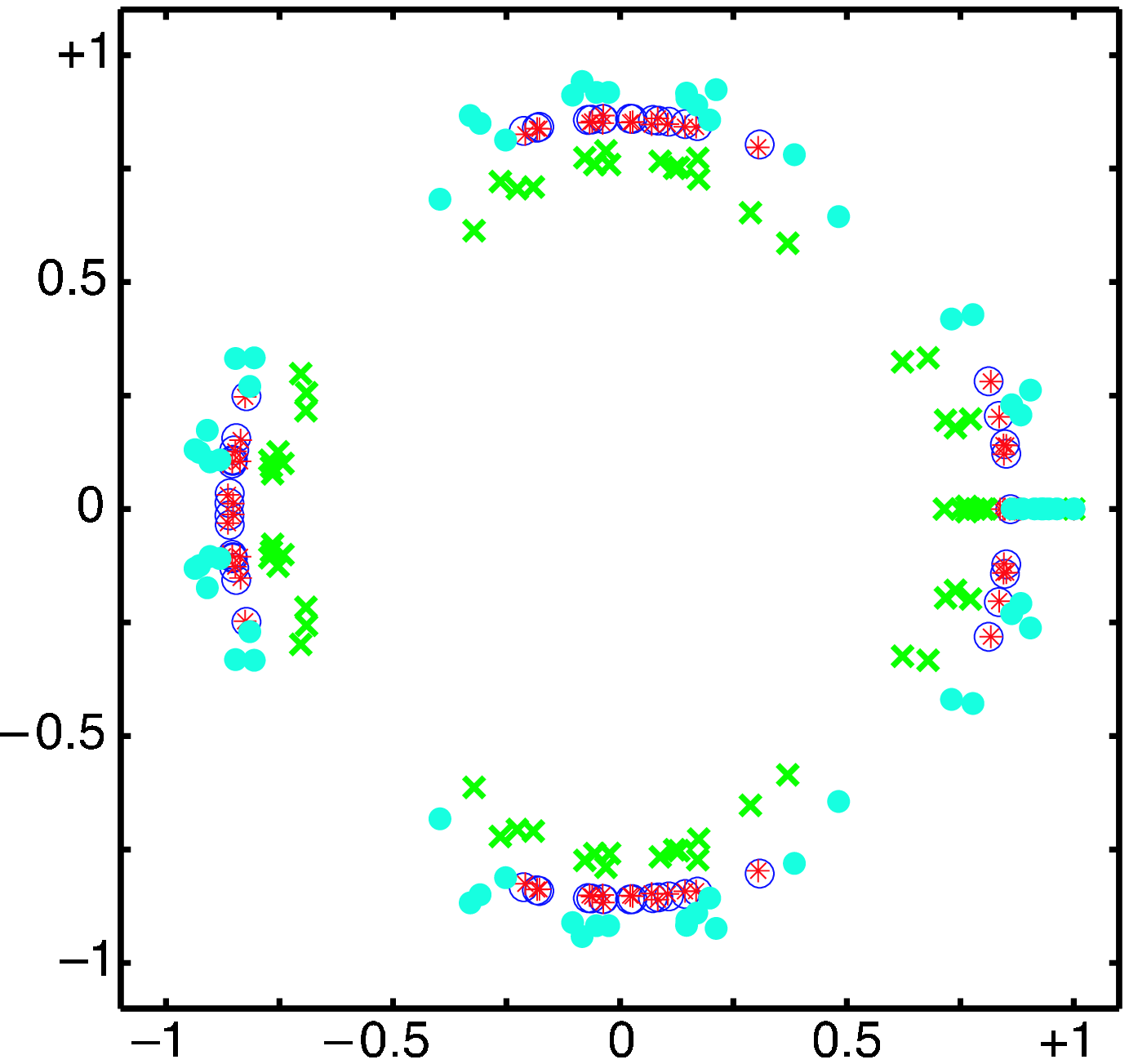}
\figcap{
Plot of the eigenvalues of simulated QFT superoperators in the complex plane,
specifically, the results of simulations with only coherent errors
(blue ``\protect\raisebox{1pt}{$\scriptscriptstyle\bigcirc$}''),
the results including relaxation using the relaxation superoperator
shown in Fig.~\ref{fig:had} and Table~\ref{tab:rel} (red ``$\,\ast\,$''),
the results including incoherent errors but no relaxation
(cyan ``{\large$\protect\bullet$}''),
and the same following simulation of the full QPT procedure (green ``$\times$''),
wherein decoherence was mimicked simply by scaling down the simulated QFT
superoperators following the state readout pulses by a factor of $0.82$ 
(save for the component with eigenvalue $1$, which is needed for trace-preservation).
To aid the comparison, the eigenvalues of the QFT superoperator
including only the coherent errors
(blue ``\protect\raisebox{1pt}{$\scriptscriptstyle\bigcirc$}'')
were scaled to have the same
RMS value as those obtained with the simulated relaxation,
thereby showing that the net effect of relaxation on the results
of the QFT is simply to lower the amplitude of the eigenvalues
without otherwise dispersing them in the complex plane.
%} \label{eigRSOdemo}
%\end{figure}
%
%\begin{figure}[H] \newpage
%\includegraphics[width=8.5cm]{eigRFIdemo}
%\renewcommand{\baselinestretch}{1.0}
%\figcap{
%The right-hand plot shows the eigenvalues of simulated QFT superoperators
%in the complex plane, first with only coherent errors as on the left,
%(blue ``\protect\raisebox{1pt}{$\scriptscriptstyle\bigcirc$}''),
%second with both coherent and incoherent errors (red $\ast$), and third
%following simulated state tomography and reconstruction (green $\times$),
%with both coherent and incoherent errors present during readout.
It may also be seen that the effect of incoherent errors on the 
eigenvalues is both to increase their angular spread and scale them
down proportionately, so that they tend to move along arcs inside
of and tangent to the unit circle, and that the additional errors
introduced by the tomography procedure do not alter them greatly.
} \label{eigRSOdemo} \label{eigRFIdemo} 
\end{figure}

We have shown that our model of the system-plus-apparatus
is able to predict many details of the experimental results
(see Table \ref{tab:allcor} for all the correlations among
the theoretical, simulated and experimental supermatrices,
together with the attenuated correlations to the theoretical).
Specifically, simulations based on the Hadamard relaxation
operator shown in Fig.~\ref{fig:had} and Table~\ref{tab:rel}
have allowed us to establish that the results contain no
specific information on the decoherence rates and processes
operative in our system, since these are averaged by the complex
sequence of transformations that make up the QFT.
Their net effect can therefore be modeled as a simple uniform attenuation
of all the product operators in the density matrix other than the identity.
This is explicitly demonstrated by the close correspondence
between the supermatrix eigenvalues shown in Fig.~\ref{eigRSOdemo},
both from simulations using the Hadamard relaxation operator
(red ``$\,\ast\,$'') as well as from simulations taking no
account of relaxation save by scaling down the nonidentity 
components of the density operators by a factor of $0.82$
(blue ``\protect\raisebox{1pt}{$\scriptscriptstyle\bigcirc$}'').
In general, of course, relaxation cannot be
accounted for by a single attenuation factor,
and in fact methods similar to those described here
have been used to determine the complete NMR relaxation
superoperator of a two-spin system \cite{Nick}.

Figure \ref{eigRFIdemo} also shows the eigenvalues of the supermatrix
obtained from simulations which included incoherent errors from RF
field inhomogeneity, first without taking relaxation or readout errors
into account (cyan ``{\large$\protect\bullet$}''), and second from
simulations which included input and output state readout and took
relaxation into account by scaling the nonidentity components of
the density matrices read out by $0.82$ (green ``$\times$'').
These plots show rather clearly that most of the additional dispersion seen
in the simulations including incoherent errors stems directly from those errors,
and was not an unintended consequence of the additional errors made during
the readout operations needed for QPT (recall that incoherent errors were
included in simulating the readout operations).

Nevertheless, the spread in the eigenvalues of the experimental
superoperator (see Fig.~\ref{eigCPTP} below) were significantly
larger than those in the simulations of the full QPT procedure,
and it was not possible to match them in a one-to-one fashion.
In part this may be due to systematic errors in the least-squares
fitting procedure by which the observables ($\mathrm m_\mathsf{xx}$
in Fig.~\ref{QPT}) were extracted from the NMR spectra, in particular
errors in distinguishing the absorptive and dispersive peak components.
Such errors are difficult to simulate accurately and therefore
not included in our simulations of the full QPT procedure,
but are consistent with the fact that the largest single-spin
corrections were $\mathsf z$-rotations (as described below).
Another, probably more important reason for these apparent
discrepancies between the model and the experiments
lies in the propagation of many small coherent errors
over the course of the QFT implementation.

Even though the simulated single-qubit gate fidelities
were all better than $0.99$, they were also less than unity.
The main thing that limited these fidelities was that the
number of parameters, and hence the number of time intervals
within which the RF amplitude and phase was held constant,
had to be kept as low as possible during the optimizations
by which the strongly modulating pulses were designed.
Although the cumulative effects of such small errors could
become significant, since they are included in the simulations
they are not likely to be the source of the additional
eigenvalue dispersion that is seen in the experimental results.
A more important source of coherent errors is expected
to be due to the fact that the protons were not included
in any of the simulations, and their couplings to the
carbons used as qubits will show up as phase rotations.
In addition, the model ignores such fine details
of the apparatus as the limits on the rise-and-fall times
of the transmitter used to generate the RF control sequences,
and more generally its frequency response characteristics.
Although the simulation of reactive circuits is
more complex than the simulation of resistive circuits,
in due course we will also include such effects in the model,
and continue to refine it until the only remaining
discrepancies between the simulated and experimental
results lie in the intrinsic measurement errors.

\begin{table}[b]
\renewcommand{\baselinestretch}{1.0}
\caption{ \label{tab:rotfix} \medskip} \parbox{0.89\textwidth}{
\centering \setlength{\baselineskip}{0.6\baselineskip}
The two triples of single-spin rotations that optimize the correlation coefficient between the best unitary approximations to the largest Kraus operators of the experimental and simulated supermatrices, when the experimental supermatrix is left or right multiplied by the unitary matrix corresponding to each one of these triples.
\bigskip} 
\begin{tabular}{||@{\hspace{12pt plus 6pt minus 6pt}}
l@{\hspace{12pt plus 6pt minus 6pt}}|@{\hspace{12pt plus 6pt minus 6pt}}
c@{\hspace{12pt plus 6pt minus 6pt}}||
*{3}{@{\hspace{24pt plus 12pt minus 12pt}}
c@{\hspace{24pt plus 12pt minus 12pt}}}|c||}
\hline\hline
Side & Spin & \multicolumn{3}{c|}{\hspace{-2em}\textsf x, \textsf y
\& \textsf z\hspace{0.6pt}-\hspace{0.4pt}Directional
Cosines of Rotation Axis} & ~Rotation Angle~ \\
\hline\hline
left & 1 & $-0.992$ & $~~0.007$ & $~~0.123$ & $36.7^\circ$ \\
left & 2 & $-0.386$ & $-0.243$ & $~~0.890$ & $~9.0^\circ$ \\
left & 3 & $~~0.703$ & $~~0.701$ & $-0.123$ & $16.9^\circ$ \\
right & 1 & $~~0.059$ & $-0.031$ & $~~0.998$ & $14.2^\circ$ \\
right & 2 & $~~0.227$ & $-0.512$ & $~~0.829$ & $10.2^\circ$ \\
right & 3 & $-0.092$ & $-0.292$ & $-0.952$ & $38.3^\circ$ \\
\hline\hline
\end{tabular} \\\bigskip
\end{table}

Unwanted bilinear interactions are refocussed during the QFT
implementation by the application of $\pi$-rotations to the
individual spins, using traditional NMR methods \cite{Ernst}.
Errors in these rotations will contribute to the bilinear errors
only to second order, so these bilinear errors depend mainly upon
the timing of the pulses, which can be controlled very accurately.
As a consequence, we expect that the residual coherent errors that were not
taken into account by the simulations will be primarily single-spin rotations.
To test this hypothesis, the error operator
$U_\Delta = U_{1,\mathsf{exp}}\, U_{1,\mathsf{sim}}^\dag$ was taken,
where $U_{1,\mathsf{exp}}$ is the best unitary approximation to the largest Kraus
operator $A_1$ of the experimental supermatrix (cf.~Fig.~\ref{ExpVsSimKrAmp}),
and $U_{1,\mathsf{sim}}$ is similarly the best unitary approximation
to the largest Kraus operator of the simulated supermatrix.
Using a numerical search, the product of three single-spin rotations
was found that fit $U_\Delta$ best in the least squares sense.
The resulting unitary $U_\Delta^1 \otimes  U_\Delta^2 \otimes U_\Delta^3$ had
a correlation coefficient with $U_\Delta$ of $0.96$, in support of the hypothesis.

\begin{figure}[H] \newpage
\includegraphics[width=3.5in]{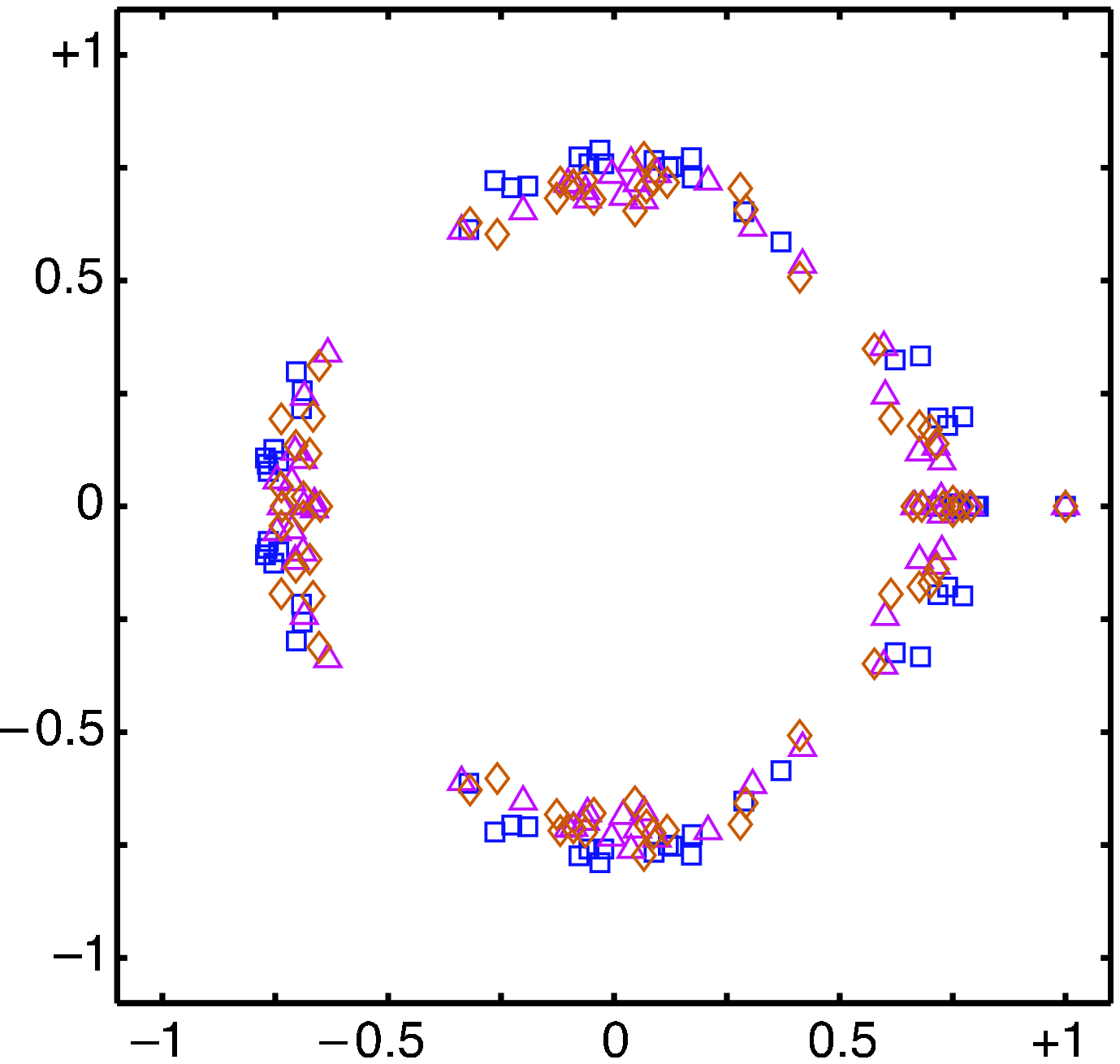}
\figcap{ 
The blue squares ($\scriptstyle\square$) are the eigenvalues
of the simulated supermatrix $\mathcal M_\mathsf{sim}$,
including coherent, incoherent and readout errors,
while the magenta triangles ($\scriptstyle\triangle$) are those
after correcting $\mathcal M_\mathsf{obs}$ by post-multiplication
with the product of single-spin rotations $U_\Delta^1 \otimes  U_\Delta^2
\otimes U_\Delta^3$ and the orange diamonds after pre-multiplication with
a product of different single-spin rotations (see Table \ref{tab:rotfix}).
Although we cannot quite unambiguously match up pairs of eigenvalues,
it is clear that they have been made very similar by these corrections.
} \label{eigSimCor}
\end{figure}

The angles and directional cosines of the axes of these three rotations
are shown in Table \ref{tab:rotfix}, while the eigenvalues of the
resulting corrected superoperators are shown in Fig.~\ref{eigSimCor}.
It should be noted that, although the correlation between $U_{1,\mathsf{exp}}$ and
$U_{1,\mathsf{sim}}$ increased from $0.90$ to $0.96$ on left-multiplying $U_1$
by the Hermitian conjugate of this product of single-spin rotation operators,
these rotations are the cumulative result of many small rotation errors
and are not simply traced back to any single short-coming in the experiments.
For completeness, the axes and angles of the single-spin rotations
that best fit the error operator $U_\Delta = U_{1,\mathsf{sim}}^\dag\,
U_{1,\mathsf{exp}}$ are also shown in Table \ref{tab:rotfix},
together with the eigenvalues of the corresponding
corrected superoperators in Fig.~\ref{eigSimCor}.
The product of these rotations similarly has a $0.97$ correlation
with $U_\Delta$, but in this case one must right-multiply
$U_{1,\mathsf{exp}}$ by the Hermitian conjugate of the product to correct it.
The close co-incidence between the angles of rotation about
the $\mathsf x$-axis on spin $1$ in the first case and about
the $\mathsf z$-axis on spin $3$ in the latter case is expected,
since $U_\mathsf{QFT} \sigma_\mathsf x^1 U_\mathsf{QFT}^\dag = \sigma_\mathsf z^3$
(recall $\sigma_\mathsf x^1\sigma_\mathsf z^3$ is a fixed point of the QFT).

\begin{table}[b]
\renewcommand{\baselinestretch}{1.0}
\caption{ \label{tab:cptp} \medskip} \parbox{0.96\textwidth}{
\centering \setlength{\baselineskip}{0.6\baselineskip}
The correlation coefficients between Theoretical, Simulated and Experimental supermatrices (see text) and the optimum completely positive and trace-preserving approximation $\mathcal M_\mathsf{CPTP}$ to the experimentally determined supermatrix $\mathcal M_\mathsf{obs}$, as well as the best unitary approximation to its largest Kraus operator $\bar U_{1,\mathsf{CPTP}} \otimes U_{1,\mathsf{CPTP}}$ and $\mathcal M_\mathsf{obs}$ itself for comparison.
\bigskip} 
\begin{tabular}{||@{\hspace{12pt plus 6pt minus 6pt}}
l@{\hspace{12pt plus 6pt minus 6pt}}||c|c|c||}
\hline\hline
Supermatrix \rule{0pt}{3.5ex} &
~\parbox{3cm}{\baselineskip=12pt Correlation with \\
$\bar U_{1,\mathsf{CPTP}} \otimes U_{1,\mathsf{CPTP}}$}~
& ~\parbox{3cm}{\baselineskip=12pt
Correlation with $\mathcal M_\mathsf{CPTP}$}~
& ~\parbox{3cm}{\baselineskip=12pt
Correlation with $\mathcal M_\mathsf{obs}$}~ \\
\hline\hline
Theoretical & $0.86$ & $0.89$ & $0.82$ \\
Simulated & $0.82$ & $0.82$ & $0.79$ \\
Experimental & $0.95$ & $0.97$ & $1.00$ \\
\hline\hline
\end{tabular} \\\bigskip
\end{table}

Finally, it is of interest to demonstrate that despite a substantial number
of negative eigenvalues in the Choi matrix of the experimental supermatrix,
it is not necessary to change it much in order to obtain a supermatrix
which represents a completely positive and trace-preserving superoperator.
For this reason the supermatrix $\mathcal M_\mathsf{CPTP}$ which best-fit 
the Choi matrix of the experimental supermatrix subject to the constraint
that it was both positive semidefinite and satisfied the trace-preservation
conditions was computed as described in Section \ref{sec:incoh}.
Although this procedure made essentially no change in the
largest Kraus operator (as expected), it did have a significant
effect on the experimental supermatrix as a whole.
The correlations between this CPTP-fit and the other supermatrices
that we have dealt with up to now are given in Table~\ref{tab:cptp},
along with those to the original experimental supermatrix for comparison.
This shows that even though imposing the complete positivity
constraint on the experimental observations did not change
the supermatrix very much, the change was distinctly in the
right direction since it improved the correlation with both
the simulated and theoretical supermatrices.
This is further confirmed by Fig.~\ref{eigCPTP}, which shows
the eigenvalues of $\mathcal M_\mathsf{CPTP}$, along with those of
the superoperator $\bar U_{1,\mathsf{CPTP}} \otimes U_{1,\mathsf{CPTP}}$
obtained from the best unitary approximation to its largest Kraus operator
(scaled down so as to have the same trace as $\mathcal M_\mathsf{CPTP}$),
and those of the experimental supermatrix $\mathcal M_\mathsf{obs}$ for comparison.
From this we see that the eigenvalues of $\mathcal M_\mathsf{CPTP}$
are closer to being cocircular, indicating that it is closer to an
attenuated unitary than was $\mathcal M_\mathsf{obs}$, but that
only after taking the unitary part did they become perfectly cocircular.

\section{Conclusions}
In conclusion, we have implemented quantum process tomography of the
quantum Fourier transform on a three-qubit NMR quantum information processor.
The overall gate fidelity (attenuated correlation between
superoperator matrices) was $0.64$, whereas the (unattenuated)
supermatrix correlation was $0.82$ (see Table \ref{tab:allcor}).
Judging by the fact that making the unitary part of the largest Kraus operator
correspond as closely as possible to the theoretical QFT gave a correlation
of $0.95$, we conclude that the loss of fidelity due to incoherence and/or
measurement errors during state tomography was of order $5$\%.
The loss of magnetization due to incoherence and decoherence,
on the other hand, was $0.64/0.82 = 0.78$, i.e.~about $22$\%.
This implies that the cummulative effects of coherent errors
reduced the fidelity by about $0.82 / 0.95 = 0.86$, or $14$\%,
consistent with the fact that the correlation between the largest Kraus
operator and the theoretical QFT unitary was $0.925 \approx \sqrt{0.86}$.

More importantly, QPT of the QFT has enabled us to validate the essential
correctness of our model of the system-plus-apparatus used for the
experiments in great detail, and to isolate its remaining shortcomings.
It has further prompted us to develop a range of data analysis
and visualization techniques for quantum process tomography,
which should be broadly applicable in quantum information processing.
The experiments described here demonstrate the precision with which
complex quantum dynamics can be controlled, and highlights the
significance of liquid-state NMR as a test bed for achieving such control.
While full QPT on larger quantum systems will never be practical,
the analysis done here should serve as an initial guide as to how
information about the errors in quantum information processors can
be extracted, and perhaps someday, how to debug a quantum computer.  

\pagebreak[4]
\bigskip\centerline{ACKNOWLEDGEMENTS}\smallskip
The authors thank J.P. Paz for helpful discussions. 
This work was supported by ARDA/ARO grants DAAD19-01-1-0519 \&
DAAD19-01-1-0654, DARPA grant MDA972-01-1-0003, NSF grant
EEC 0085557, the Air Force Office of Sponsored Research,
and by the Cambridge-MIT Institute, Ltd.

%\end{twocolumn}

\newpage
\mylistoffigs

\begin{figure}[H] \newpage
\includegraphics[width=7cm]{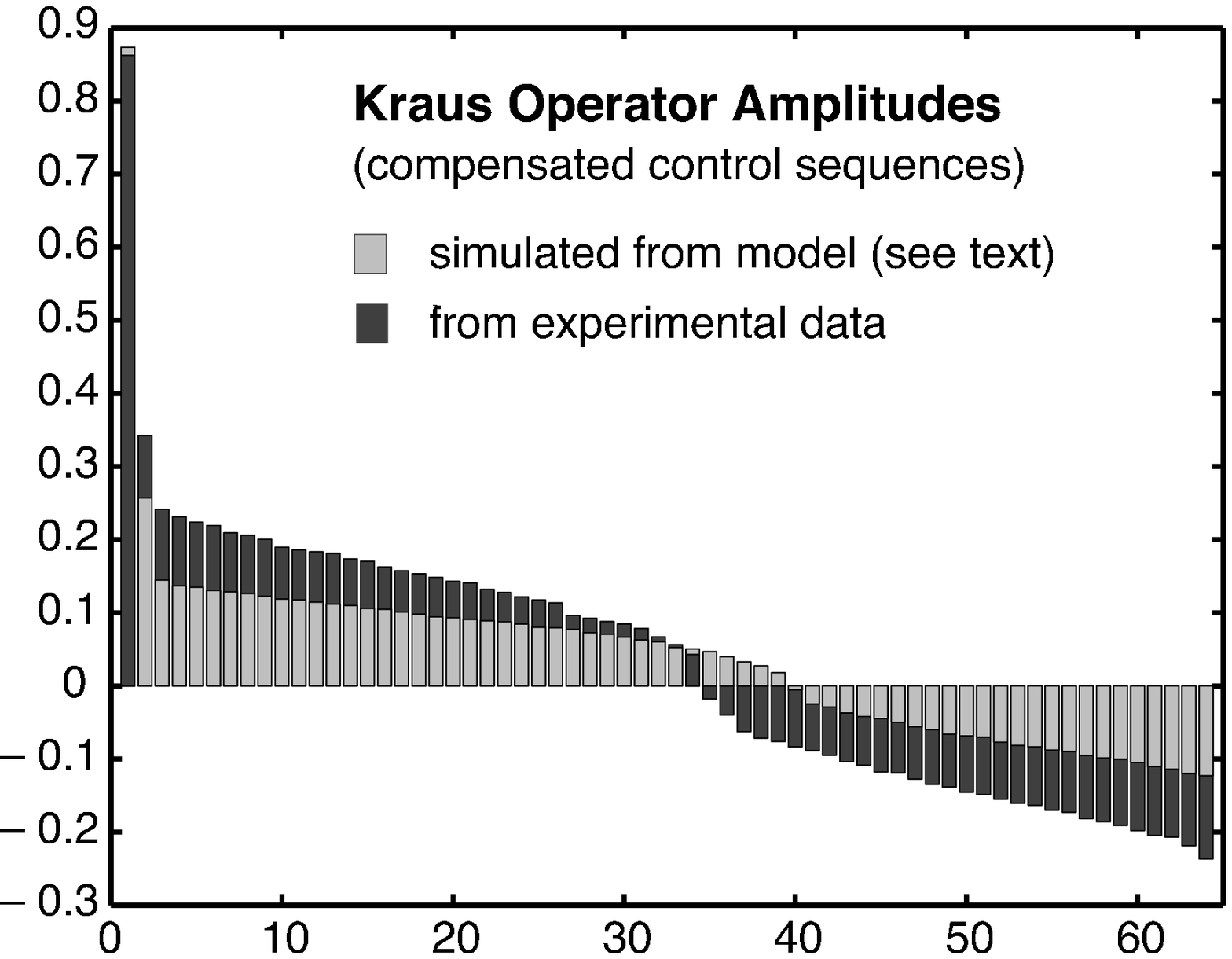} \hspace{0.5cm}
\includegraphics[width=8.5cm,height=6.5cm]{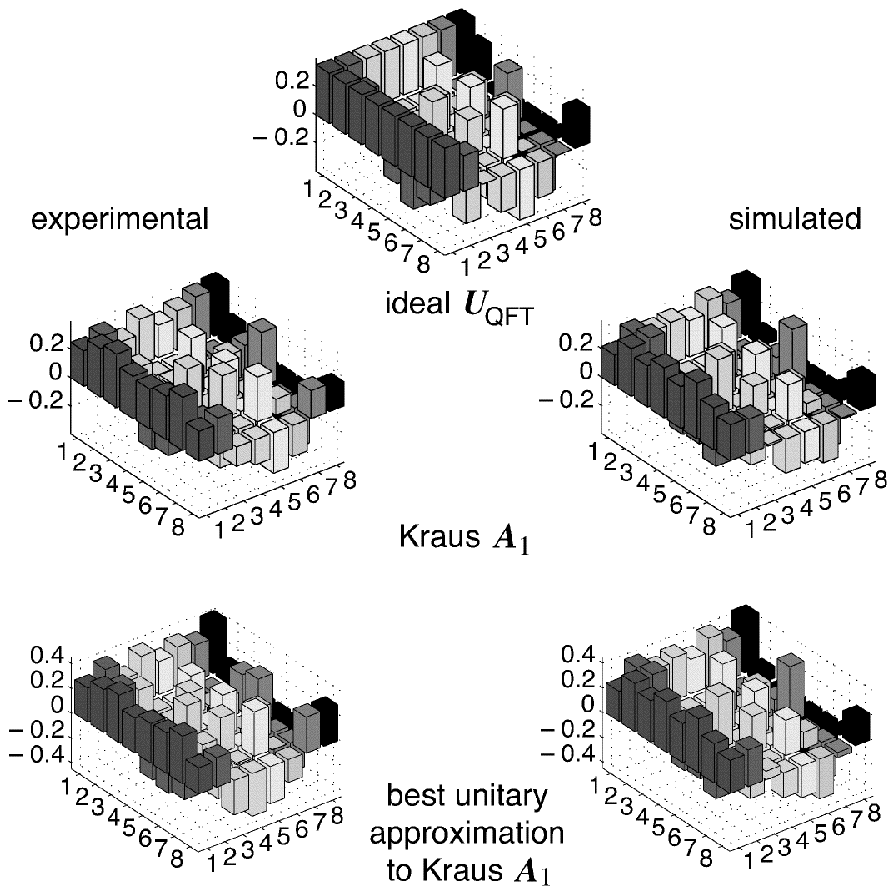}
\renewcommand{\baselinestretch}{1.0}
\figcap{ The left-hand plot shows the real amplitudes, 
of experimental QFT Kraus operators (dark grey) and those obtained from
the QPT simulation based on the system-plus-apparatus model (light grey),
where the negative values plotted are actually the negative square roots of
one-eighth the absolute values of the corresponding Choi matrix eigenvalues.
It may be seen that the experimental deviated significantly more
from being completely positive than the simulated, most likely to the
absence of errors from fitting the spectra in the latter (see text).
The dominant Kraus operator has an amplitude $0.86$ for the experiment
and $0.87$ for the model. The right side of the figure compares the real part of
the theoretical unitary QFT operator (top-middle) to the largest Kraus operator
from the experiment (middle-left) and simulated (middle-right) supermatrices.
Also shown are the real parts of the corresponding best unitary approximations
to these largest Kraus operators (bottom line). The correlation between
the two largest Kraus operators, and between the corresponding best
unitary approximations, was $0.90$ in both cases.
} \label{ExpVsSimKrAmp}
\end{figure}

\pagebreak[4]

\begin{figure}[H] \newpage
\includegraphics[height=4.75cm]{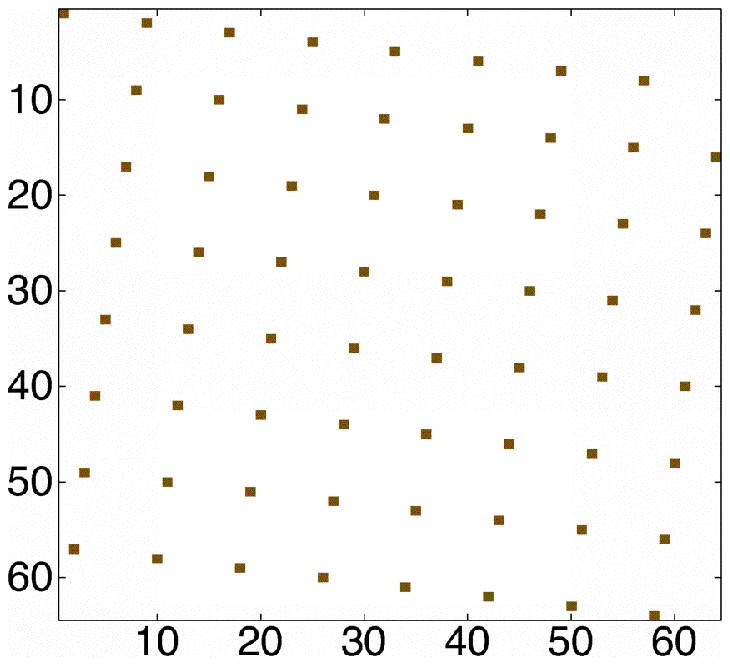}
\includegraphics[height=4.75cm]{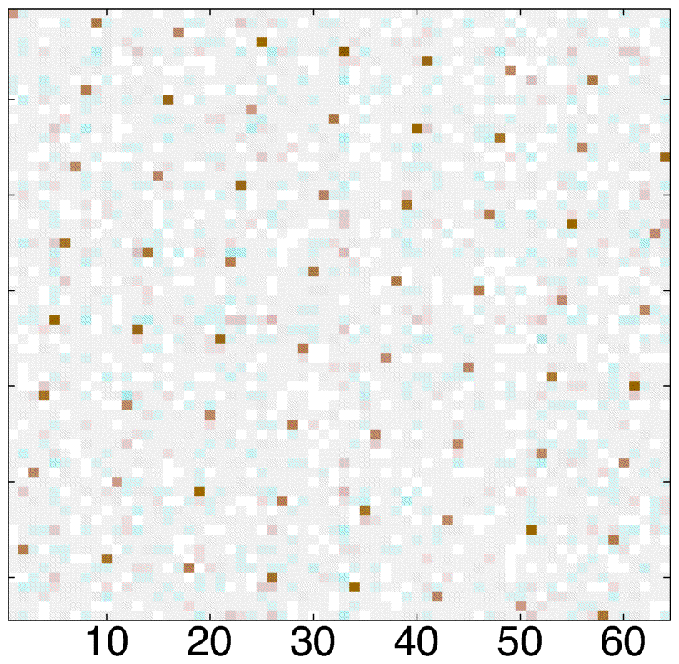}
\includegraphics[height=4.85cm]{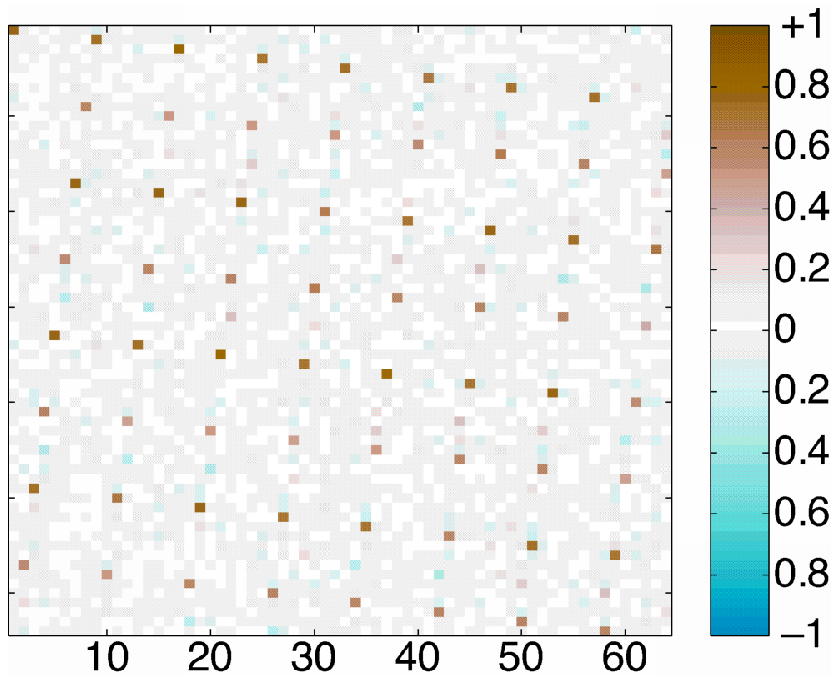}
\renewcommand{\baselinestretch}{1.0}
\figcap{
The theoretical (left), simulated (middle) and experimental
(right)  supermatrices of the QFT in the phase space basis,
wherein all their elements are necessarily real.
Since the theoretical is simply a permutation matrix in this basis,
the differences between it and the other two are easily seen,
but the physical meanings of these differences are obscure.
For example, the level of ``background noise'' is noticeably
larger in the simulation than in the experiment, although
both have the maximum entry of each row (column)
in the same place as the theoretical.
} \label{PhSp}
\end{figure}

\begin{figure}[H] \newpage
\includegraphics[height=4.75cm]{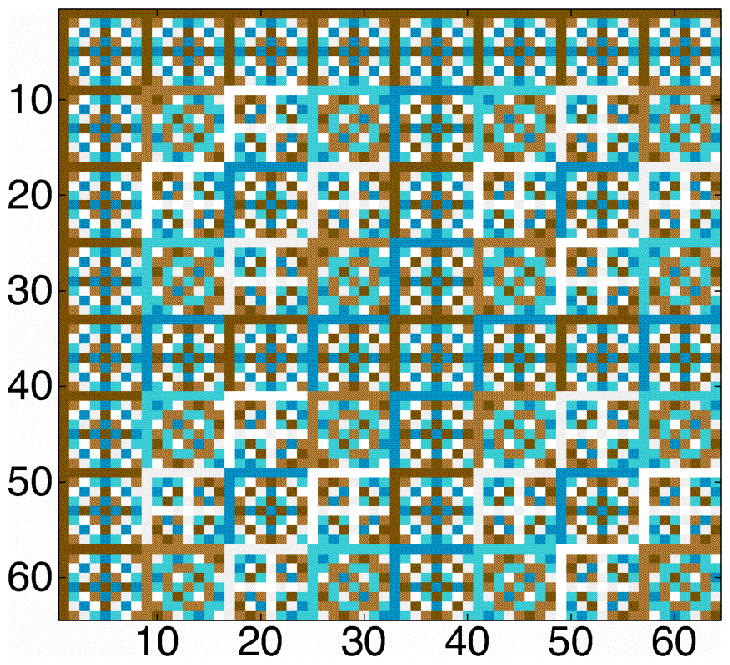}
\includegraphics[height=4.75cm]{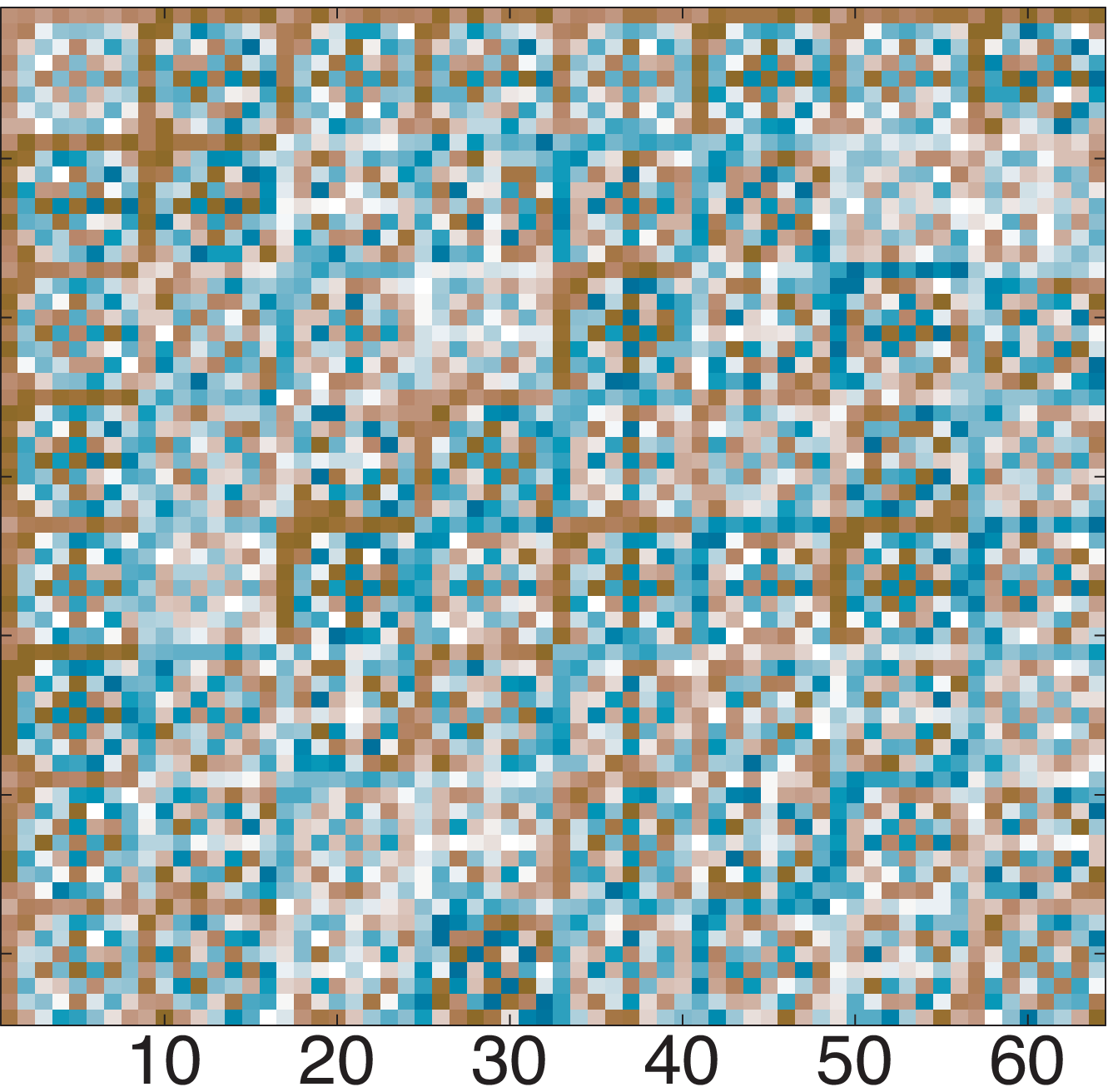}
\includegraphics[height=4.9cm]{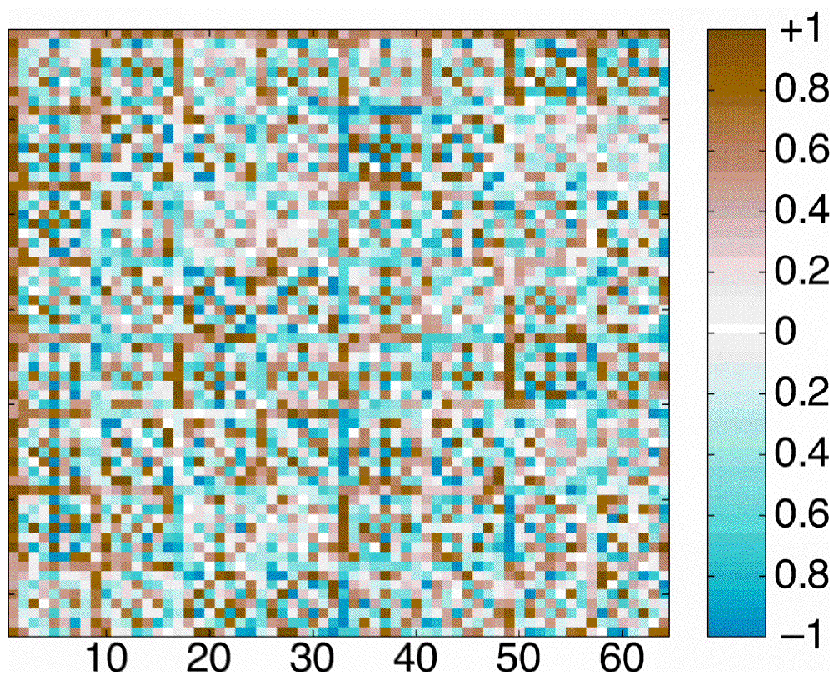}
\renewcommand{\baselinestretch}{1.0}
\figcap{
The real parts of the theoretical (left), simulated (middle) and
experimental (right) supermatrices in the computational (Zeeman) basis,
where the theoretical superoperator is given in terms of the unitary operator
by $\overline{U}_\mathsf{th}^\mathsf{QFT}\otimes U_\mathsf{th}^\mathsf{QFT}$.
The correlations and attenuated correlations among
these matrices may be found in Table \ref{tab:allcor}. 
Although the pattern of the theoretical superoperator's matrix elements
is visible in both the simulated and the experimental supermatrix, 
it is still not at all obvious from this vantage point what errors 
have occurred during implementation or how they can be corrected.
} \label{Zeem}
\end{figure}

\begin{figure}[H] \newpage
\includegraphics[height=4.75cm]{TheQFTSupPSB}
\includegraphics[height=4.75cm]{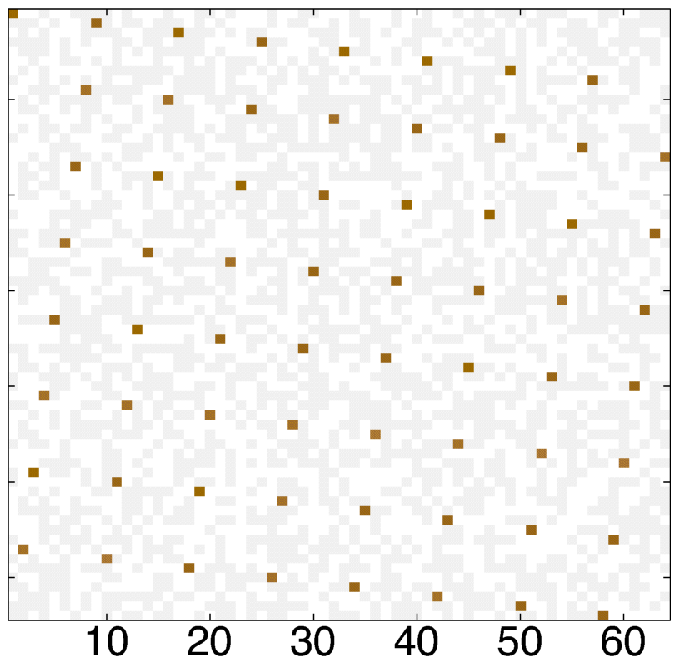}
\includegraphics[height=4.85cm]{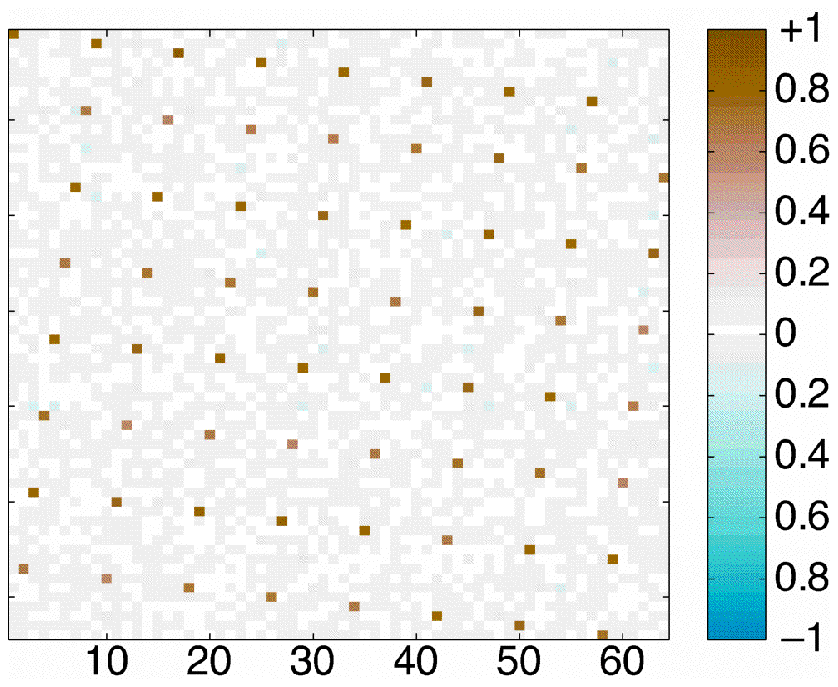}
\figcap{
Theoretical (left), simulated (middle) and experimental (right) supermatrices
in the phase space basis after correcting the latter two by a unitary
supermatrix designed to bring them as close as possible to the theoretical
(as described in the text). The levels of background noise are a great deal
less than in the corresponding uncorrected supermatrices (Fig.~\ref{PhSp}).
} \label{CorrectedSuperPSB}
\end{figure}

\begin{figure}[H] \newpage
\includegraphics[width=3.5in]{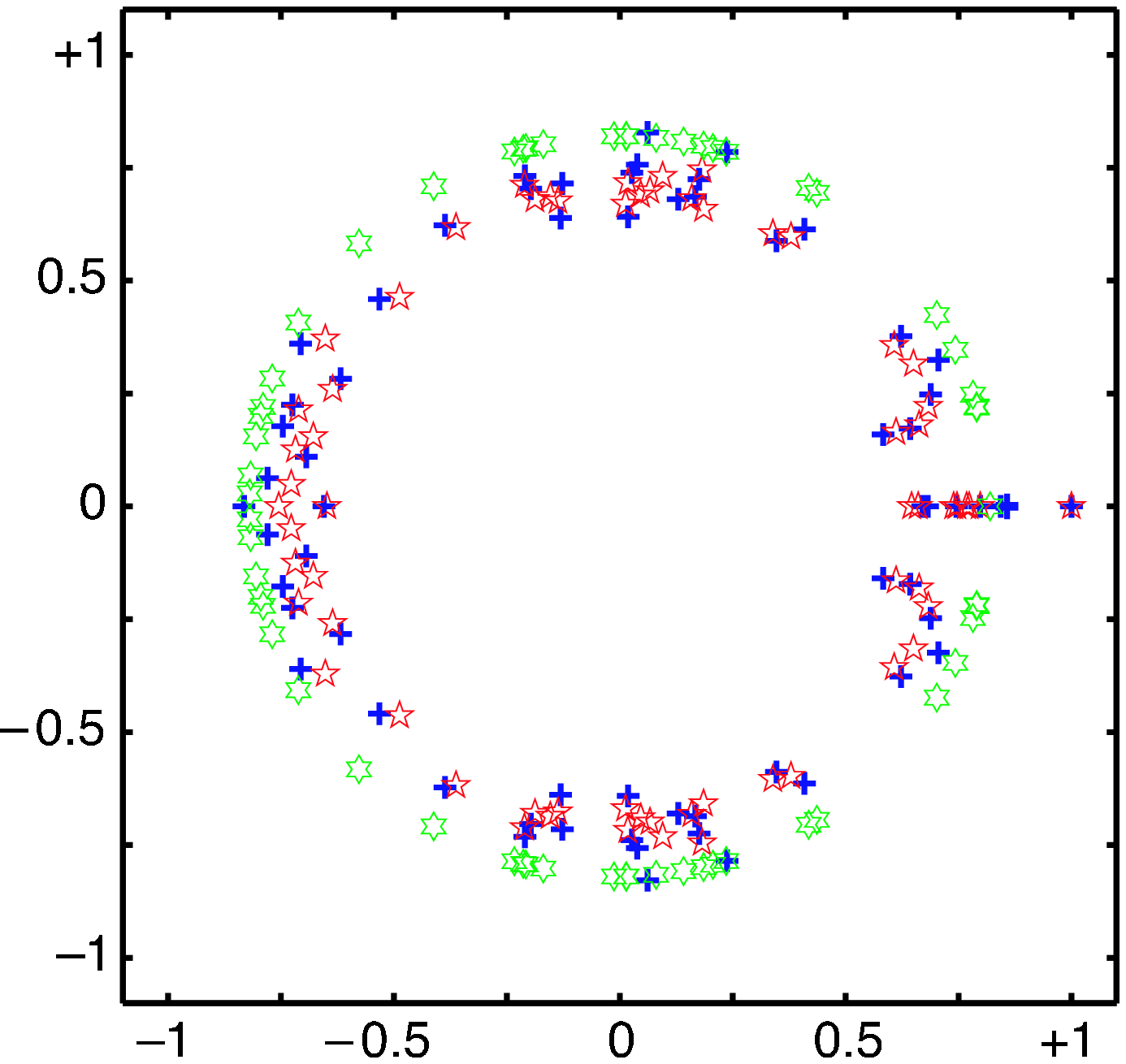}
\figcap{
The blue plus-signs
(\protect\rule[1.5pt]{7pt}{1pt}\hspace{-4pt}\protect\rule[-1.5pt]{1pt}{7pt}~)
are the eigenvalues of the experimental supermatrix $\mathcal M_\mathsf{obs}$,
while the red five-poined stars (\protect\raisebox{1pt}{$\scriptstyle\bigstar$})
are those its best completely positive and trace-preserving 
approximation $\mathcal M_\mathsf{CPTP}$, and the green six-pointed stars
(\protect\raisebox{1pt}{$\blacktriangle$}\hspace{-0.72em}\protect\raisebox{-1pt}{$\blacktriangledown$})
are those of the superoperator corresponding to the
best unitary approximation to its largest Kraus operator
$\bar U_{1,\mathsf{CPTP}} \otimes U_{1,\mathsf{CPTP}}$ (see text),
scaled down to have the same trace as $\mathcal M_\mathsf{CPTP}$.
} \label{eigCPTP} 
\end{figure}


\begin{thebibliography}{99}
\bibitem{Schu}
B. Schumacher, Phys. Rev. A, {\bf 54}, 2614, (1996).
\bibitem{Grum}
G. Teklemariam, E. M. Fortunato, M. A. Pravia, T. F. Havel, and D. G. Cory, 
Phys. Rev. Lett. {\bf 86}, 5845 (2001).
\bibitem{E}
E. M. Fortunato, M. A. Pravia, N. Boulant, G. Teklemariam, T. F. Havel, D. G.
Cory, J. Chem. Phys. {\bf 116}, 7599-7606, (2002).
\bibitem{Swap}
N. Boulant, K. Edmonds, J. Yang, M. A. Pravia, and D. G. Cory, Phys. Rev. A
\textbf{68}, 032305 (2003).
\bibitem{C+N}
I. L. Chuang, M. A. Nielsen, J. Mod. Optics {\bf 44}, 2455 (1997).
\bibitem{PCZ}
J. F. Poyatos, J. I.Cirac, P. Zoller, Phys. Rev. Lett. {\bf 78}, 390 (1997).
\bibitem{DM}
G. M. D'Ariano, L. Maccone, Phys. Rev. Lett. {\bf 80}, 5465 (1998);
G. M. D'Ariano, M. G. A. Paris, M. F. Sacchi, {\it Advances in Imaging and Electron Physics} (Academic Press, Englewood Cliffs, NJ), in press, {\tt http:arxiv.org/abs/quant-ph/0302028}.
\bibitem{YSW}
Y. S. Weinstein, M. A. Pravia, E. M. Fortunato, S. Lloyd, D. G. Cory, 
Phys. Rev. Lett. {\bf 86}, 1889 (2001).
\bibitem{CP}
M. H. Levitt, R. Freeman, J. Magn. Reson. {\bf 33}, 473 (1979);
%R. Freeman, S. P. Kempsell, M. H. Levitt, J. Magn. Reson. {\bf 38}, 453 (1980);
M. Levitt, J.  Magn. Reson. {\bf 48}, 234 (1982); R. Tycko, Phys. Rev. Lett. {\bf 51}, 775 
(1983); A. Shaka, R. Freeman, J. Magn. Reson., {\bf 55}, 487 (1983); M. Levitt,
Prog. Nucl. Magn. Reson. Spect., {\bf 18}, 61 (1986).
\bibitem{Marco}
M. A. Pravia, N. Boulant, J. Emerson, A. Farid, E. M. Fortunato, T. F. Havel,
R. Martinez, D. G. Cory, J. Chem. Phys. {\bf 119}, 9003 (2003). 
\bibitem{K}
K. Kraus, {\it States, Effects, and Operations} (Springer-Verlag, Berlin, 
FRG, 1983).
\bibitem{Leung}
D. W. Leung, J. Math. Phys. {\bf 44}, 528 (2003).
\bibitem{Tim}
T. F. Havel, J. Math. Phys. {\bf 44}, 534 (2003).
\bibitem{Andrew}
A. M. Childs, I. L. Chuang, D. W. Leung, Phys. Rev. A {\bf 64}, 012314, (2001). 
\bibitem{Nick}
N. Boulant, T. F. Havel, M. A. Pravia, D. G. Cory, Phys. Rev. A {\bf 67}, 042322 (2003).
\bibitem{Shor} 
P. W. Shor, Siam. J. Comput. {\bf 26} 1484-1509 (1997).
\bibitem{Josza}
R. Jozsa, Proc. Roy. Soc. Lond. {\bf 454} (1998).
\bibitem{Zalka}
C. Zalka, Proc. Roy. Soc. Lond. A {\bf 454} 313-322, (1998); 
S. Wiesner, quant-ph/9603028.
\bibitem{Sch}
R. Schack, Phys. Rev. A {\bf 57} (1998).
\bibitem{Cop}
D. Coppersmith, IBM Research Report RC19642 (1994).
\bibitem{Price}
M. D. Price, T. F. Havel and D. G. Cory, New J. Phys. {\bf 2}, 10 (2000).
\bibitem{Baker}
Y. S. Weinstein, S. Lloyd, J. Emerson, D. G. Cory, Phys. Rev. Lett. 
{\bf 89}, 157902 (2002).
\bibitem{Tim2}
T.F. Havel, Quant. Inf. Proc., {\bf 1}, 511, (2003).
\bibitem{P}
P. Pechukas, Phys. Rev. Lett. {\bf 73}, 8, (1994).
\bibitem{S1}
P. Stelmachovic, B. Buzek, Phys. Rev. A {\bf 64}, 062106, (2001).
\bibitem{S2}
P. Stelmachovic, B. Buzek, Phys. Rev. A {\bf 67}, 029902, (2003).
\bibitem{N}
N. Boulant, S. Furuta, J. Emerson, T. F. Havel, D. G. Cory, quant-ph/0312116.
%\bibitem{Schu2}
%B. Schumacher, M.A. Nielsen, Phys. Rev A {\bf 54}, 2629 (1996). 
%\bibitem{QCC}
%S. Lloyd, Phys. Rev. A {\bf 55}, 1613 (1997). 
\bibitem{Cory1}
D. G. Cory, R. Laflamme, E. Knill, L. Viola, T. F. Havel, N. Boulant,
G. Boutis, E. Fortunato, S. Lloyd, R. Martinez, C. Negrevergne, M. Pravia,
Y. Sharf, G. Teklemarian, Y. S. Weinstein, Z. H. Zurek, Fortsch. Phys. {\bf 48},
875 (2000).
\bibitem{Jones}
J. A. Jones, Fortsch. Phys. {\bf 48}, 325 (2001).
\bibitem{Shy}
S. S. Somaroo, D. G. Cory, T. F. Havel, Phys. Lett. A, {\bf 240}, (1998).
\bibitem{Tim3}
T.F. Havel, Y. Sharf, L. Viola, D.G. Cory, Phys. Lett. A, {\bf 280}, 282 
(2001).
\bibitem{Ernst}
R. R. Ernst, G. Bodenhausen, A. Wokaun, {\it Principles of Nuclear Magnetic 
Resonance in One and Two Dimensions}, Oxford Univ. Press, Oxford (1994).
\bibitem{CPS}
C. Miquel, J.P. Paz, M. Saraceno, Phys. Rev. A, {\bf 65}, 062309, (2002).
%\bibitem{C}
%D. G. Cory, M. D. Price, W. Mass, E. Knill, R. Laflamme, W. H. Zurek, T. F. 
%Havel, S. S. Somaroo, Phys. Rev. Lett. {\bf 81}, 2152-2155, (1998).

\end{thebibliography}
\end{document}